\documentclass[final,5p,times,times]{elsarticle}
\biboptions{sort&compress}
\usepackage{hyperref}
\usepackage{amssymb}
\usepackage{graphicx}
\usepackage{xcolor,amsmath,amsfonts,graphicx,amssymb,subfigure}
\usepackage[utf8]{inputenc}
\usepackage{mathtools}
\usepackage{natbib}
\usepackage{graphicx}
\usepackage{amsmath}
\usepackage{slashed}
\usepackage{multirow}
\usepackage{amssymb,amsmath,amsfonts}

\graphicspath{{figures/}}

\graphicspath{{./figures/}}
\DeclareMathOperator{\tr}{tr}

\graphicspath{{./figures/}}

\begin{document}

\begin{frontmatter}
\title{QCD transition at the physical point,\\and its scaling window from
twisted mass Wilson  fermions}
\author[Juelich,Moscow1]{Andrey Yu. Kotov}
\ead{a.kotov@fz-juelich.de}
\author[Florence]{Maria Paola Lombardo}
\ead{lombardo@fi.infn.it}
\author[Samara]{Anton Trunin}
\ead{amtrnn@gmail.com}
\address[Juelich]{Juelich Supercomputing Centre, Forschungszentrum Juelich, D-52428 Juelich, Germany}
\address[Moscow1]{Bogoliubov Laboratory of Theoretical Physics, Joint Institute for Nuclear Research, Dubna, 141980 Russia}
\address[Florence]{INFN, Sezione di Firenze, 50019 Sesto Fiorentino (FI), Italy}
\address[Samara]{Samara National Research University, Samara, 443086 Russia}
\begin{abstract}
We study the scaling properties of the finite temperature QCD phase transition, for light quark masses
ranging from the heavy quark regime to their physical values.
The lattice results are obtained
in the fixed scale approach from  simulations of $N_f=2+1+1$ flavours of Wilson fermions at maximal twist. 
We identify an order parameter free from the the linear contributions in mass  due to  additive renormalization and 
regular terms in the Equation of State, which proves useful for the assessment of the hypothesized universal behaviour. We find 
compatibility with the 3D $O(4)$ universality class for the physical pion mass  
and temperatures $120$~MeV $ \lesssim  T \lesssim   300$~MeV. We discuss violation of scaling at larger masses 
and a possible cross-over to mean field behaviour. 
The chiral extrapolation $T_0 = 134^{+6}_{-4}$~MeV of the pseudocritical  temperature is robust against predictions  of different universality classes and consistent with its estimate from the  $O(4)$ Equation of State for the  physical pion mass.
\end{abstract}
\end{frontmatter}

\section{Introduction}

The appearance of pseudo-Goldstone bosons in the spectrum signals the spontaneous breaking of chiral symmetry of strong interactions. It is well known that chiral symmetry is restored at high temperatures, and imprinting of this phenomenon for physical masses is visible in the behaviour of the order parameters and in the spectrum. 
If the symmetry restoring transition is continuous, the link between the genuine critical behaviour in the chiral limit and the observations at finite masses is made transparent by the universal Equation of State (EoS).
This is true only within a limited region 
around criticality -- the scaling window. 

Despite substantial progress ~\cite{Ding:2020rtq,Guenther:2021lrv}, the nature of the continuum limit and the extent of its scaling window  are still open issues. Chiral and axial symmetries play a pivotal role here: 
in this study we make use, as in our previous work, 
of Wilson fermions at maximal twist \cite{Kotov:2020hzm,Kotov:2019dby,Burger:2018fvb,Burger:2017xkz,Burger:2011zc}, 
a lattice formulation with good chiral properties and an alternative
to the more used staggered fermions. 

The global symmetry of the QCD Lagrangian 
$U(n)_L\times U(n)_R \cong SU(n)\times SU(n)\!\times\! U(1)_V\times U(1)_A$,
valid at classical level, is broken by topological quantum fluctuations.
In the limit of an infinite strange mass 
there are at least two possible scenarios for the high temperature
transition depending
on the fate of the axial symmetry~\cite{Pisarski:1983ms,Rajagopal:1992qz,Pelissetto:2013hqa}: if the axial symmetry breaking is not much sensitive
to the chiral restoration, the breaking pattern is 
$SU(2)_L\!\times\!SU(2)_R \to SU(2)_V$.
Due to the associate diverging correlation length, the theory is effectively 3D, in the universality class
of the classical four-component Heisenberg antiferromagnet: $O(4) \to O(3)$ \cite{Pisarski:1983ms}. 
If instead axial symmetry is correlated with chiral symmetry, the relevant breaking pattern would be 
$U(2)_L\!\times\!U(2)_R \to U(2)_V$, hinting either at a first 
or at a second order transition with different exponents~\cite{Pelissetto:2013hqa}. 
 
We are interested in properties of the transition for physical
values of the strange quark mass $m_s$.  If the two flavor
transition is of first order, we are likely dealing with a first order transition for any $m_s$. Alternatively, if the $m_s=0$ first order transition ends at $m_s = m_\text{crit}^s$,   $m_s$ merely renormalizes the coefficients of the effective action for $m_s > m_\text{crit}^s$~\cite{Pisarski:1983ms,Rajagopal:1992qz},   without altering  the critical behaviour. In this case one conventionally assumes that there is a line of second
order transition  $T_c(m_l=0, m_s) > T_c(m_l=0, m_s) $  for $m_l=0$, $\infty > m_s > m_\text{crit}^s$. 
Away from the true critical point dimensional reduction may fail, resulting in a 4D theory and a mean field behaviour. The extent of the scaling window
and the threshold of dimension reduction are non-universal features, which should be investigated by ab-initio methods.

These important issues are addressed by
model studies~\cite{Kawarabayashi:1980dp,Kawarabayashi:1980uh,Nicola:2019ohb,Horvatic:2018ztu},
phenomenological analysis~\cite{Kapusta:2019ktm,Kapusta:1995ww,Shuryak:1993ee}, 
Functional Renormalization Group~\cite{Resch:2017vjs,Schaefer:2013isa,Gao:2020qsj,Braun:2020ada}, and mostly on the lattice~\cite{Mazur:2018pjw,Sharma:2018syt,Fukaya:2017wfq,Sharma:2017yjc,Schmidt:2017bjt,Tomiya:2016jwr,Aarts:2019hrg,Aarts:2020vyb,Kotov:2019dby,Kotov:2020hzm}. 
Consistency with $O(4)$ scaling has been reported in lattice studies of the theory with two light and strange
flavors, at low pion mass~\cite{Ding:2019prx,Ding:2020rtq}, once different sources of scaling violations
are taken into account.
The same study finds a critical temperature in the chiral limit $T_0 = 132^{+3}_{-6}$~MeV.
Consistency with $O(4)$ scaling was found also by earlier studies with Wilson fermions, see e.g. \cite{Burger:2011zc,Umeda:2016qdo,Ejiri:2009ac}. Concerning axial symmetry, 
 the current understanding is that 
it seems to be effectively restored above $T_c$ ~\cite{Ding:2020xlj,Kaczmarek:2021ser,Kaczmarek:2020sif, Aoki:2021qws,Aoki:2020noz,Mazur:2018pjw,Buchoff:2013nra,Suzuki:2020rla,Kanazawa:2015xna,Aoki:2012yj,Tomiya:2016jwr,Brandt:2019ksy,Brandt:2016daq,Cossu:2013uua,Chiu:2013wwa}, but there is no consensus on the restoration temperature. On the analytic side, a 4D 
analysis~\cite{Braun:2010vd} reported  scaling only for very light pion masses, $m_\pi < 1$ MeV, 
and an apparent scaling for larger masses. The same study underscores the lack of dimensional reduction as a potential
source of scaling violations. A recent work~\cite{Braun:2020ada} confirms these findings, 
suggesting that pion masses as light as $30$ MeV
are needed to reach the scaling window in QCD, with a consistent extrapolation to $T_0 \simeq 142$ MeV
in the chiral limit from different prescriptions. A recent review contains a short introduction to
the scaling window in QCD, and summarises these and other studies \cite{Kotov:2021hri}.

In this study we define an ad-hoc  order parameter,  
which we dub $\langle\bar\psi\psi\rangle_3$,  free from contributions linear
in mass \cite{Unger:2010wcq}. We explore a range of masses and temperatures, in an attempt to identify the limits of the scaling
window, and its possible cross-over to a mean field behaviour. A very preliminary account of some of the results
has been presented in Ref.~\cite{Kotov:2020hzm}.

\section{Observables, magnetic Equation of State,  and scaling} 

A quantitative way to describe a critical system  with a breaking field $h$ relies
on  the EoS for the order parameter $M$
\begin{equation}
    M = h^{1/\delta}f(t/h^{1/{\beta \delta}}).
    \label{eq:eos}
\end{equation}
In the Eq.~\eqref{eq:eos}  we 
identify  $M \equiv \langle \bar \psi \psi \rangle$,   $h \equiv m_l$,  
$t \equiv  (T - T_0)$, where  $m_l$ is the light quark mass
and $T_0\equiv T_c(m_l \to 0)$ is the critical temperature in the chiral limit,
$\delta$ and $\beta$ are critical exponents. 
Note that there are two independent  normalizations for $M$
and for $t$, which, like $T_0$, depend on the strange quark mass.
In our range of masses it is legitimate to trade $m_l \simeq m_{\pi}^2$,
with a suitable adjustment of normalizations. 
$f(x)$, with $x=t/h^{1/{\beta \delta}}$, is a universal curve for a given breaking pattern. 

\begin{figure}[thb]
\includegraphics[width=9cm]{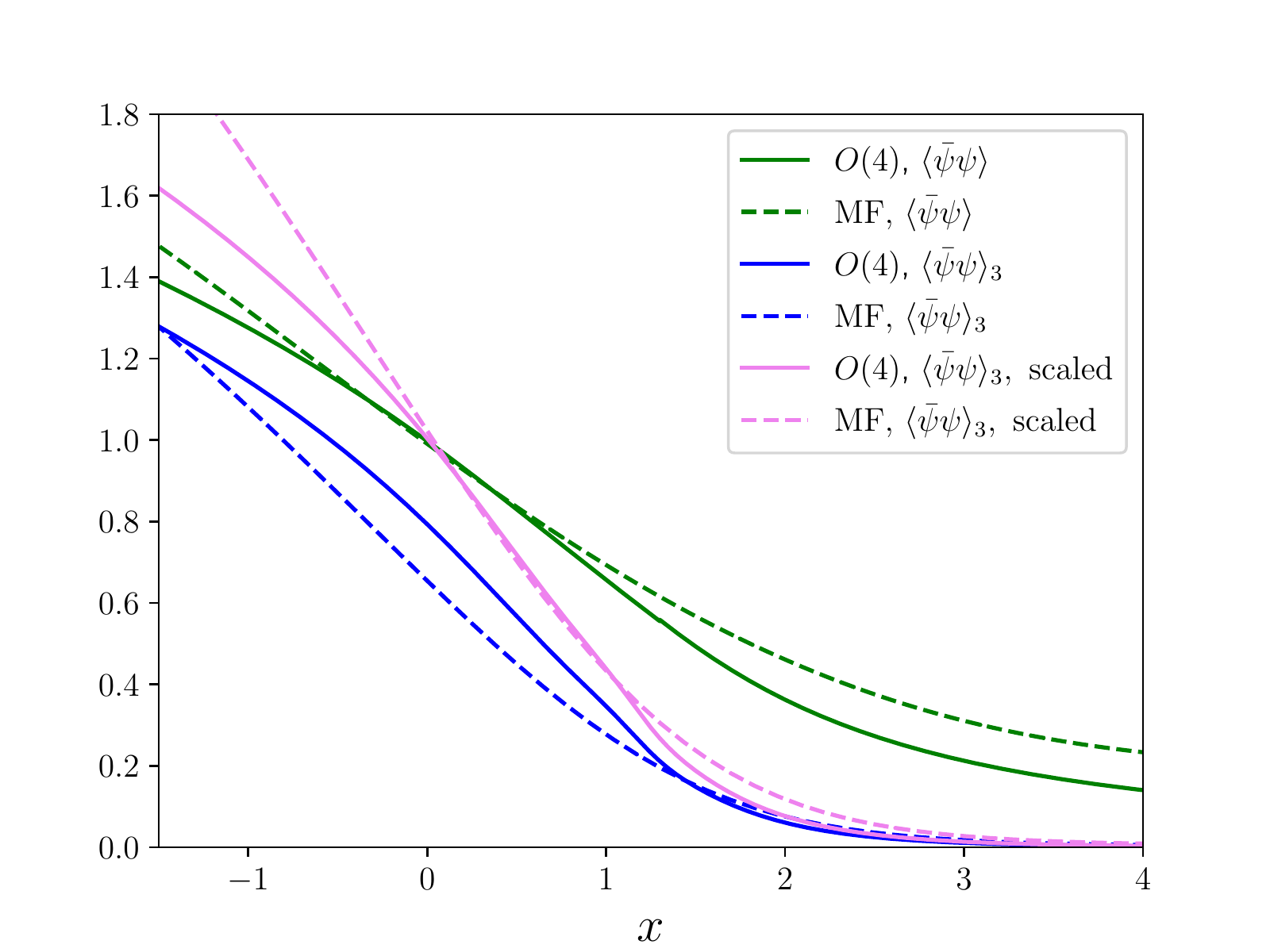}
\caption{The Equation of State for the chiral condensate $\langle\bar\psi\psi\rangle$ and for the new order parameter $\langle\bar\psi\psi\rangle_3$ in the critical region for the 3D $O(4)$ universality class, and for the mean field. For a more direct comparison we also plot the results,  normalised so that they coincide at $x=0$,  as purple lines.}
\label{fig:O4_pbp}
\end{figure}

For the 3D $O(4)$  scaling function we rely on the parameterization from Ref.~\cite{Engels:1999wf}
as well as on its polynomial interpolation~\cite{Engels:2011km,Engels:2003nq},
which is smooth in the critical region. A successful scaling of results for a finite mass would identify the critical temperature in the chiral limit, besides confirming universality. 
One less stringent approach to scaling relies on pseudo-critical temperatures associated with features of the order parameter. 
Consider the two susceptibilities:  $\chi_L = \frac{\partial \langle\bar \psi \psi\rangle}{\partial m_l}$
and    $\chi_\Delta =  \frac{\partial \langle\bar \psi \psi\rangle}{\partial T}$, 
which peak  at 
$t/h^{1/\beta \delta} = 1.35(3)$  
and $t/h^{1/\beta \delta} = 0.74(4)$, respectively,  
for 3D $O(4)$ universality class with the
critical exponents $\delta = 4.8(1)$ and $\beta = 0 38(1)$. 
The peak positions  define  pseudo-critical temperatures

\begin{equation}
T_c(m_\pi)= T_0 + A z_p  m_\pi^{2/\beta \delta}.
\label{eq:tcmpi}
\end{equation}
$A$ is a mass independent parameter, and $T_c$ for different observables  should scale with the same exponent 
${2/\beta \delta}$,  but with different ${z_p}'s$.

\subsection{A new order parameter}
In the same spirit as Ref. \cite{Kocic:1992is}, we 
consider the  transverse 
$\chi_T = \frac{\langle\bar \psi \psi\rangle}{m_l}$ and longitudinal 
$\chi_L = \frac{\partial \langle\bar\psi\psi\rangle}{\partial m_l}$ susceptibilities
and their difference $O_{LT} \equiv \chi_T - \chi_L$. 
$O_{LT}$ is an order parameter for the transition, since 
$\chi_L = \chi_T$ in the chiral limit in the symmetric phase. 
Moreover, by construction, any linear contribution in $m_l$ -- either
coming from the regular part of the EoS or from additive renormalization --
drops from the difference. However
$O_{LT}$ is divergent in the chiral limit in the broken phase. 
The singularity is avoided by defining: 
\begin{equation}
\langle\bar\psi\psi\rangle_3 \equiv \langle \bar \psi \psi \rangle  - m_l \,\chi_L \equiv \langle \bar \psi \psi \rangle - m_l  \frac{\partial \langle \bar\psi\psi \rangle}{\partial m_l}.
\label{eq:pbp3}
\end{equation}
This is the order parameter, whose Taylor expansion in $m_l$  -- when defined -- starts at a third order, as noted in Ref.~\cite{Unger:2010wcq}.

\begin{figure}[htb]
    \begin{center}
    \includegraphics[width=9cm]{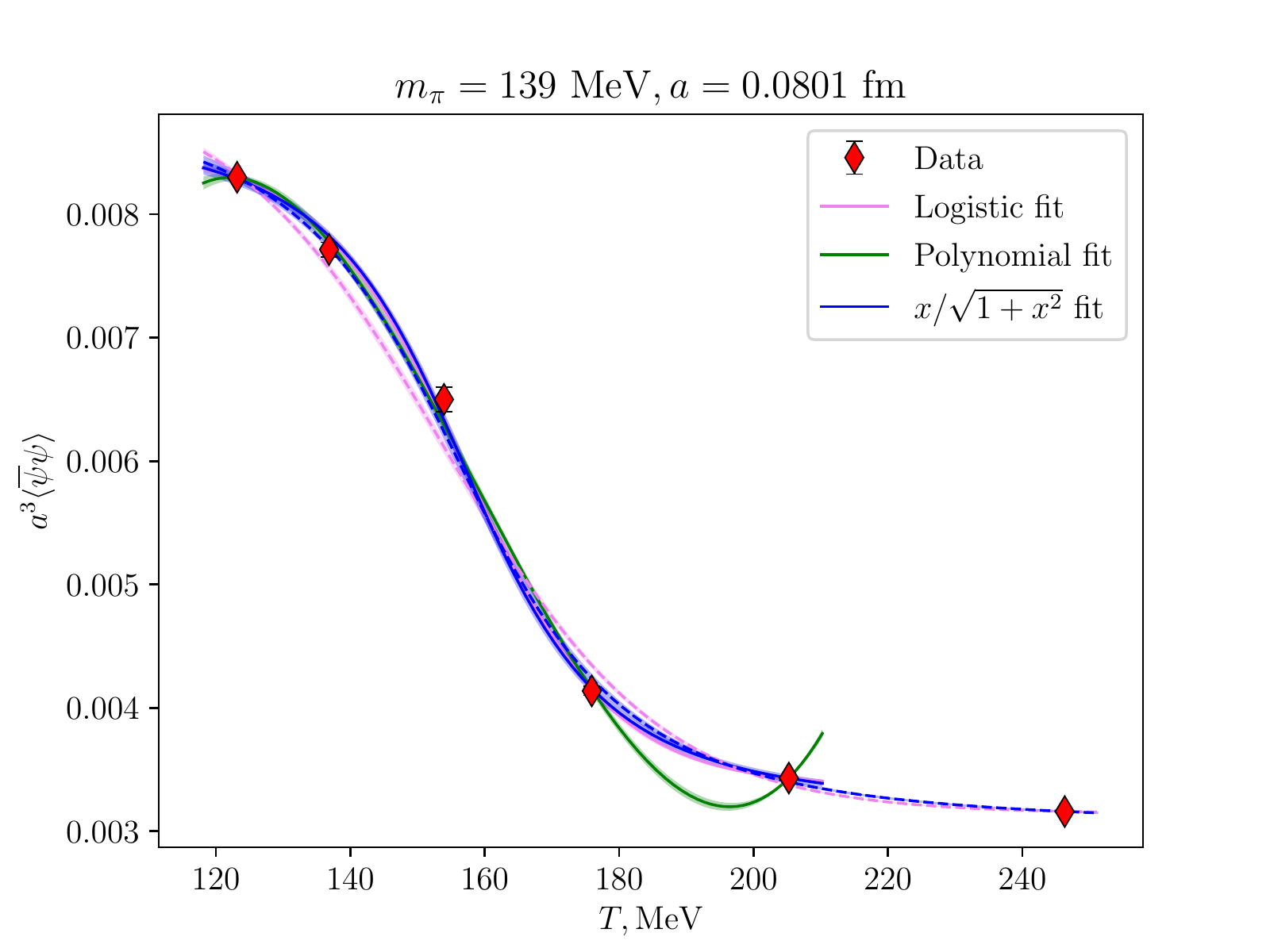}
    \caption{Chiral condensate (\ref{eq:condensate}) as a function of temperature $T$. Solid lines indicate the fit in the range $[120:210]$ MeV, dashed lines correspond to the range $[120:250]$ MeV. 
    }
    \label{fig:condensate}
    \end{center}
\end{figure}

\begin{figure*}[htb]
    \centering
    \includegraphics[width=8cm]{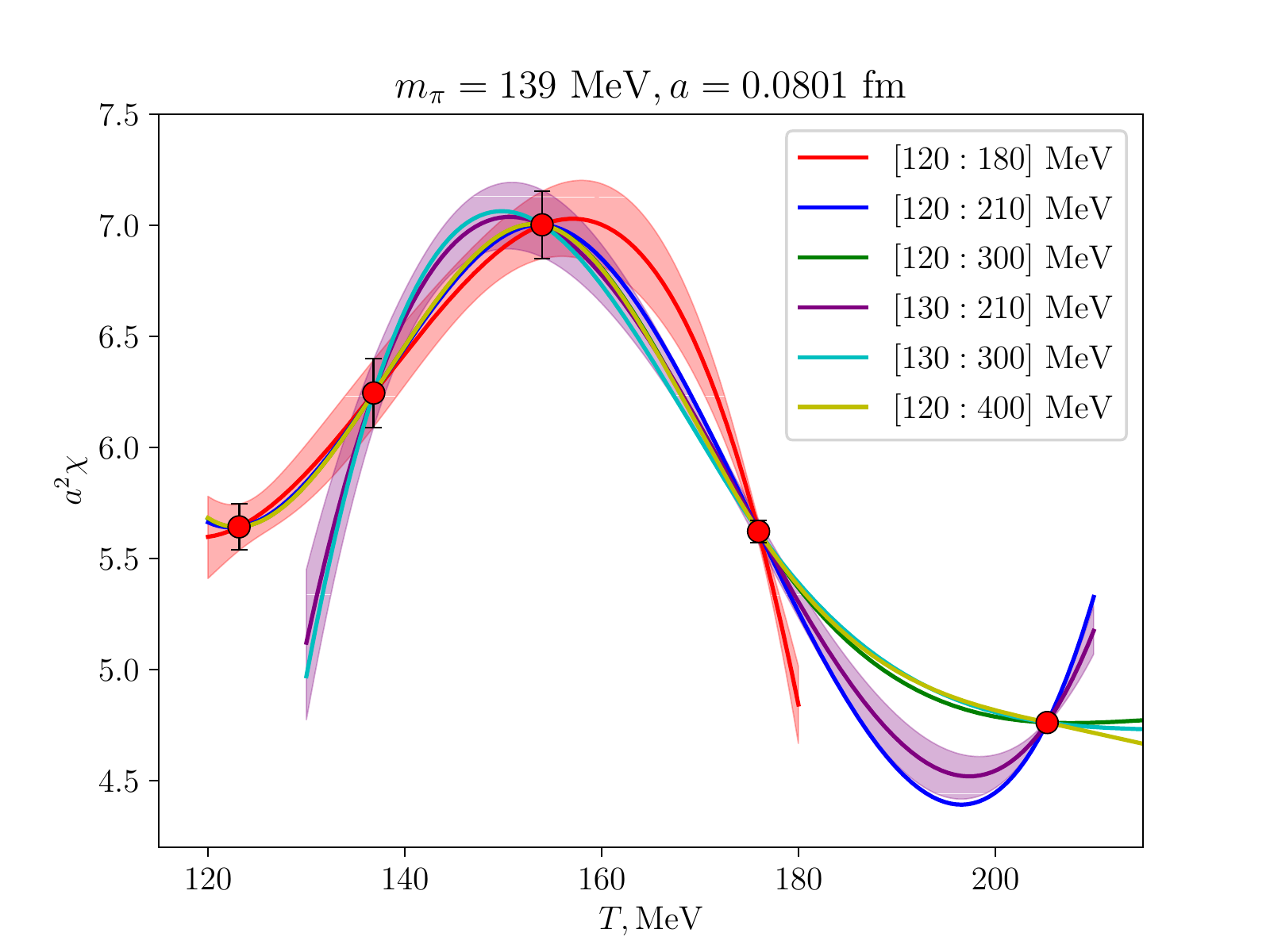}
    \includegraphics[width=8cm]{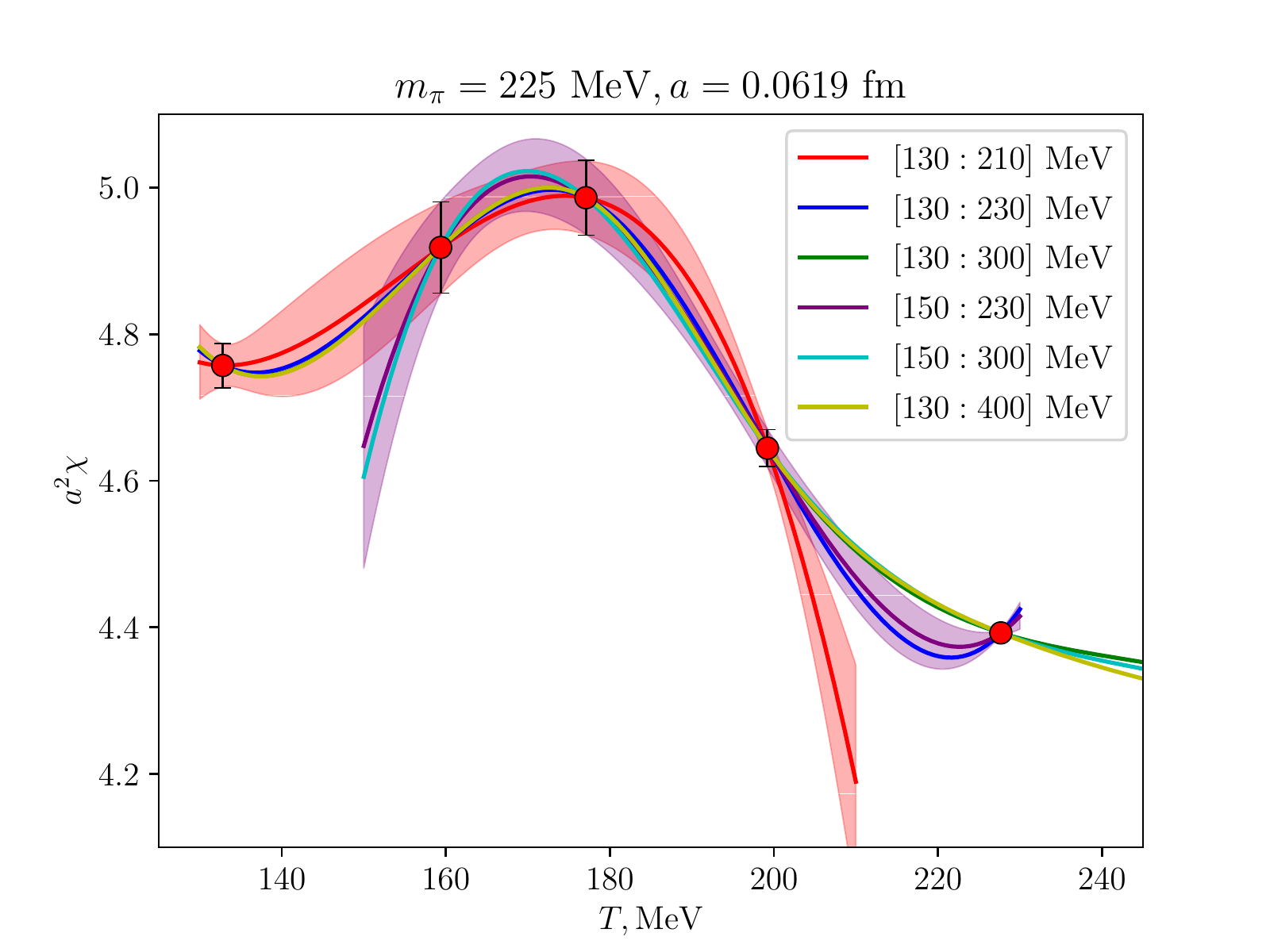}
    \includegraphics[width=8cm]{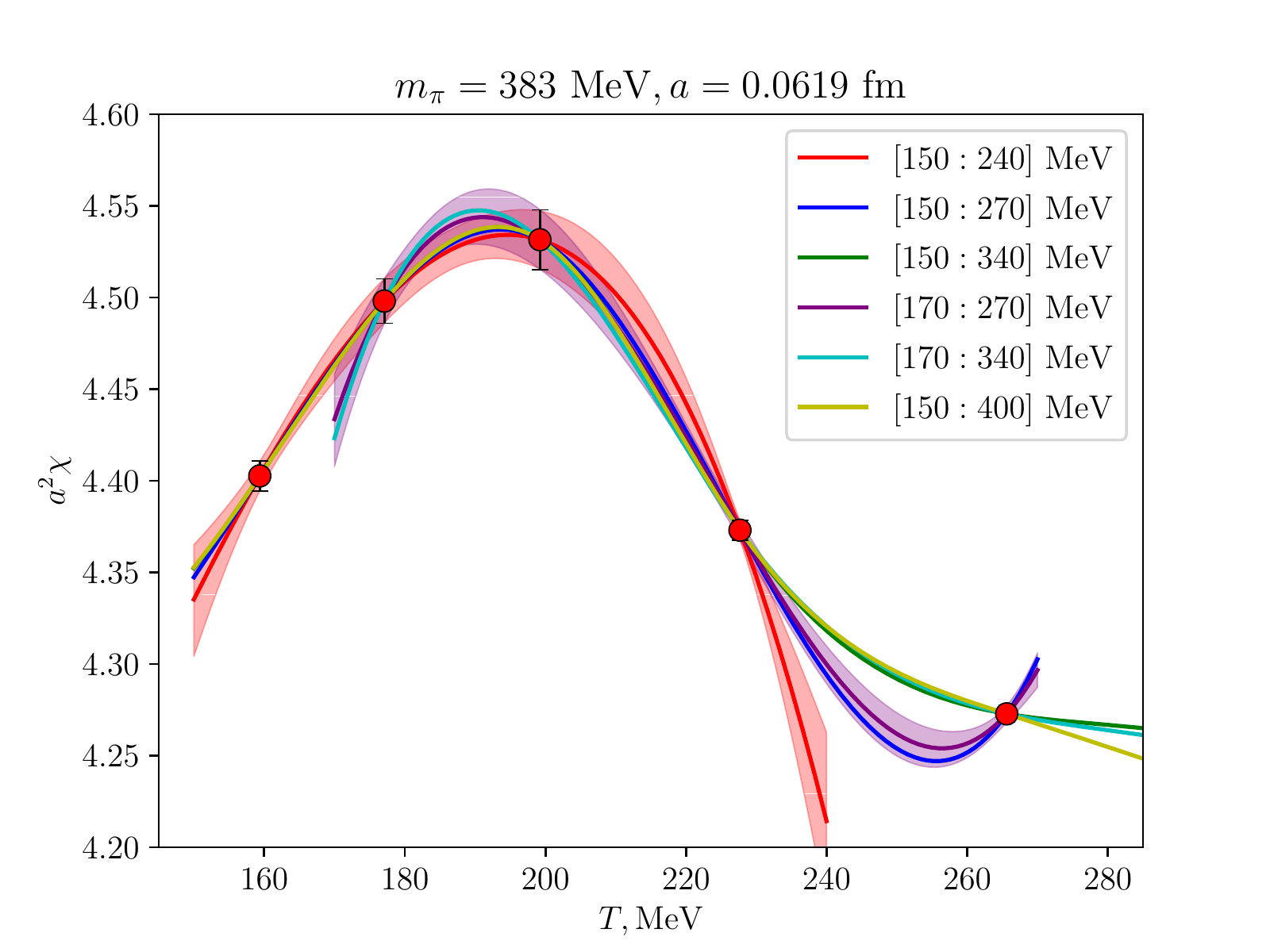}
    \includegraphics[width=8cm]{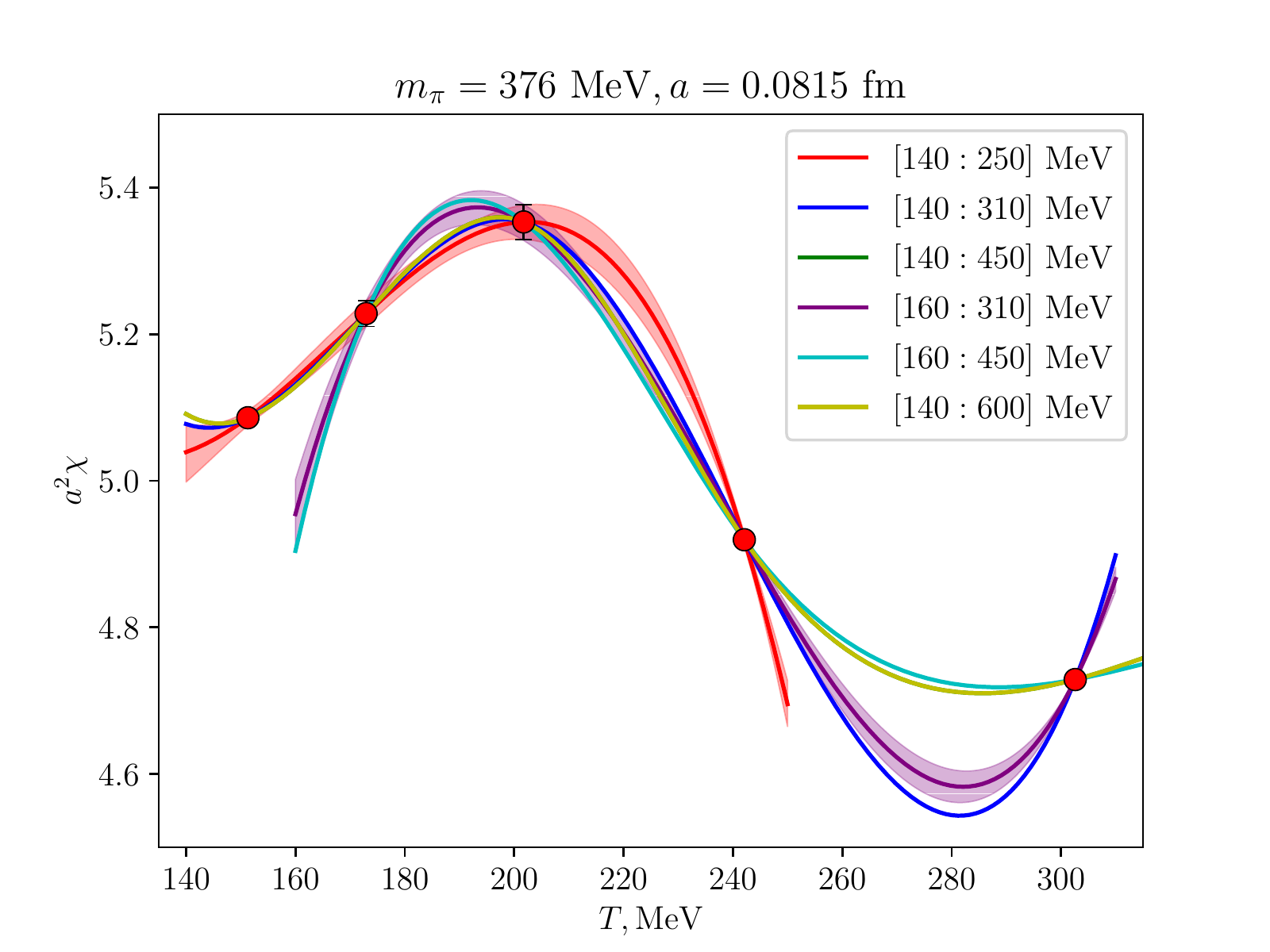}
    \caption{Chiral susceptibility as a function of temperature and its cubic spline interpolation. In some cases we also display the statistical errors, see the text.}
    \label{fig:sus}
\end{figure*}

Deriving the Equation of State for ${\langle\bar\psi\psi\rangle_3}$ is straightforward:
\begin{equation}
\frac{\langle\bar\psi\psi\rangle_3}{m_l^{1/\delta}} = f(x)( 1 - 1/\delta) + \frac{x}{\beta \delta} f(x)' .
\label{eq:eosthree}
\end{equation}
The high temperature
leading term is  
  $\langle\bar\psi\psi\rangle_3 \propto t^{-\gamma - 2 \beta \delta}$
rather than $\langle \bar\psi\psi \rangle \propto  t^{-\gamma}$: the decay is rather fast. The inflection point that drives the behaviour
of the pseudo-critical temperature associated with $\langle\bar\psi\psi\rangle_3$ is $x_{\mathrm{infl}} = 0.55(1)$.
Note that the pseudo-critical temperature  for this observable follows the same scaling in Eq.~(\ref{eq:tcmpi}), but with a smaller $z_p$: this means that the pseudo-critical temperature associated with the inflection point of 
$\langle\bar\psi\psi\rangle_3$ is smaller than the pseudo-critical temperatures associated with chiral condensate and susceptibility. This also implies that it is closer to the true critical one in the limit $m_l\to0$. 

In Fig.~\ref{fig:O4_pbp} 
we compare the EoS for $\langle\bar\psi\psi\rangle_3$ with the one for 
$\langle\bar \psi \psi\rangle$
for the $O(4)$ Universality class and for the mean field. We note
the sharper decrease of $\langle\bar \psi\psi\rangle_3$, very understandable
given that it is  closer to the chiral condensate in the chiral limit, followed by the high temperature behaviour
just described.  For either observable we also show the mean field result: it is indeed
very close to the 3D $O(4)$, so the transition from the scaling window to a regime
with small fluctuations is expected to be very smooth.

\begin{table}[thb]
\begin{center}
\begin{tabular}{|c|c|c|c|c|}
\hline
TMFT& ETMC  & $m_\pi$ [MeV] & $a$ [fm] & $Z_P$ \\
\hline
M140 & cB211.072.64 & 139.3(7) & 0.0801(4) & 0.462(4) \\
D210 & D15.48 & 225(5) & 0.0619(18) & 0.516(2)  \\
D370 & D45.32 & 383(11) & 0.0619(18) &  0.516(2) \\
B370 & B55.32 & 376(14) & 0.0815(30) &  0.509(4) \\
\hline
\end{tabular}
\caption{Parameters of the $N_f = 2 + 1 +1 $ gauge fields ensembles used for the analysis. Lattice spacing and pion mass for D and B ensembles are taken from ~\cite{Werner:2019hxc}, renormalization factor $Z_p$ for these ensembles was measured in \cite{Carrasco:2014cwa}. All parameters of the M ensemble are from \cite{Alexandrou:2020okk}. The strange and charm quark masses are close to their physical values.}
\label{table:sim}
\end{center}
\end{table}

\section{Numerical results}

\begin{table}[htb]
\begin{center}
\begin{tabular}{|c|c|c||c|c|c|}
\hline
$N_t$ & $T$ [MeV] & \# conf & $N_t$ & $T$ [MeV] & \# conf \\
\hline
20 & 123(1) & 782 & 10 & 246(1) & 592 \\
18 & 137(1) & 892 & 8 & 308(2) & 498  \\
16 & 154(1) & 534 & 6 & 411(2) & 195 \\
14 & 176(1) & 359 & 4 & 616(3) & 472  \\
12 & 205(1) & 337 &  & &  \\

\hline
\end{tabular}
\caption{Statistics of the physical pion mass ensembles M140. Each fourth molecular dynamics trajectory was saved.}
\label{tab:stat}
\end{center}
\end{table}

\begin{figure*}[thb]
    \centering
    \includegraphics[width=8cm]{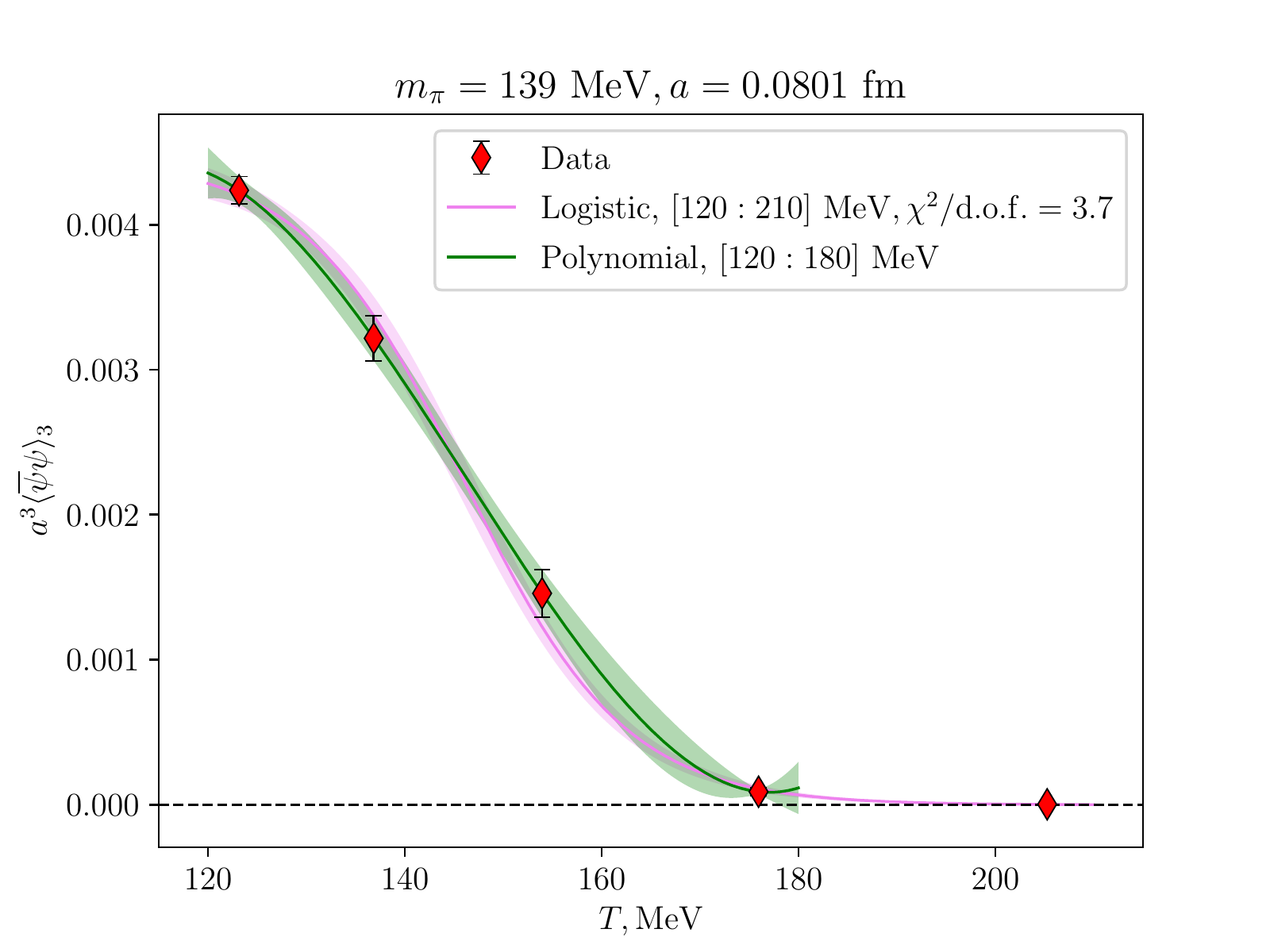}
    \includegraphics[width=8cm]{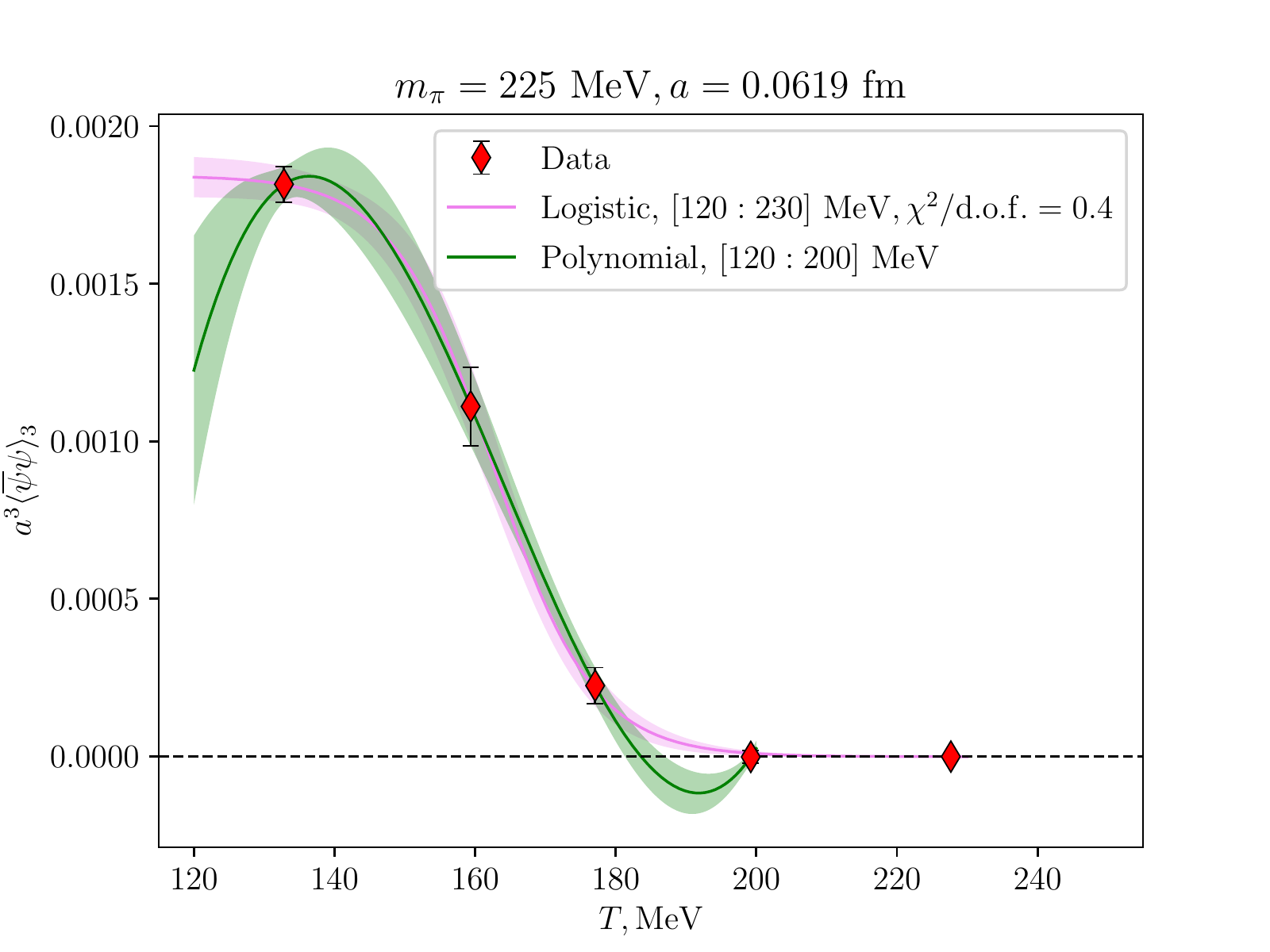}
    \includegraphics[width=8cm]{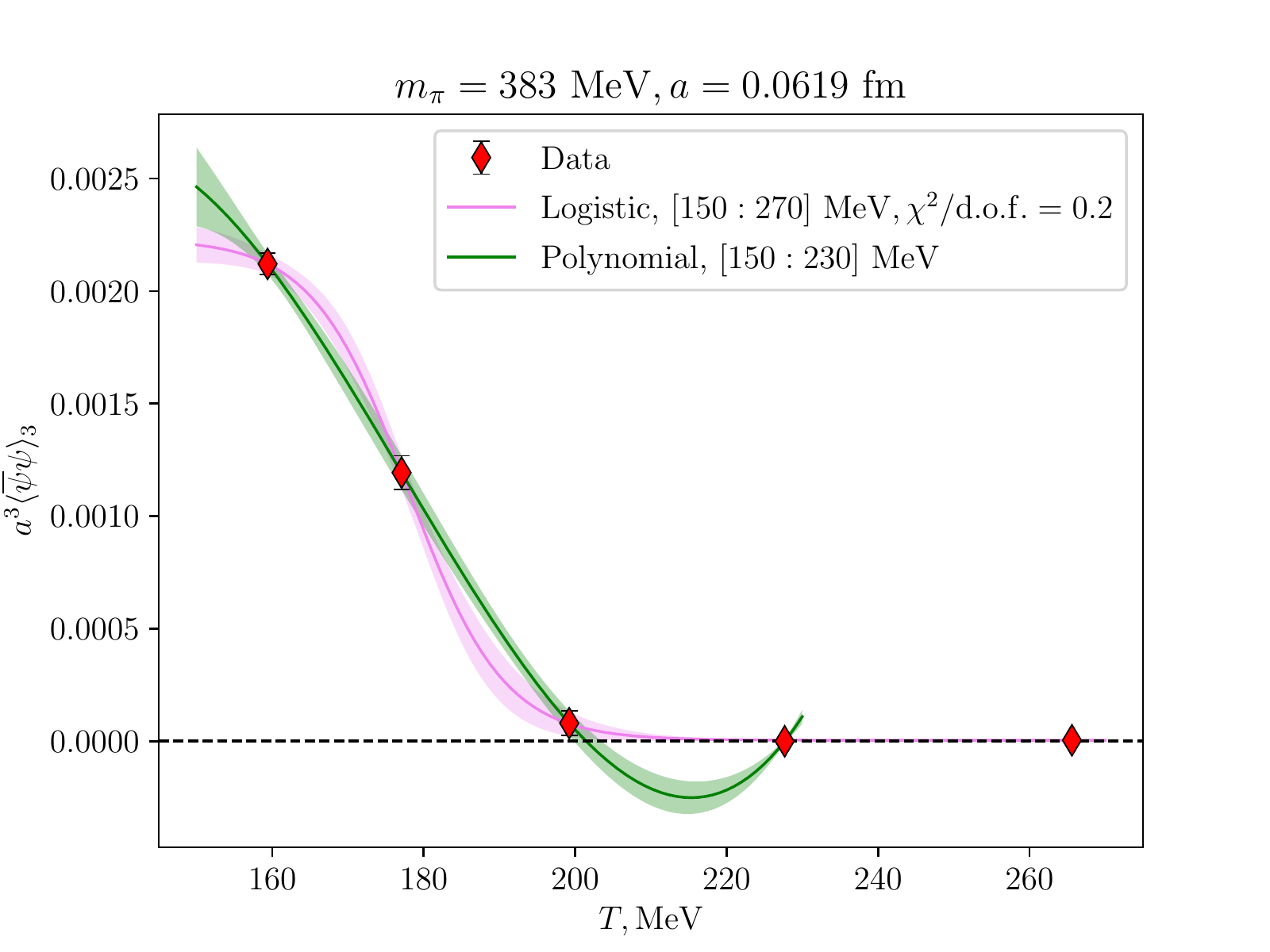}
    \includegraphics[width=8cm]{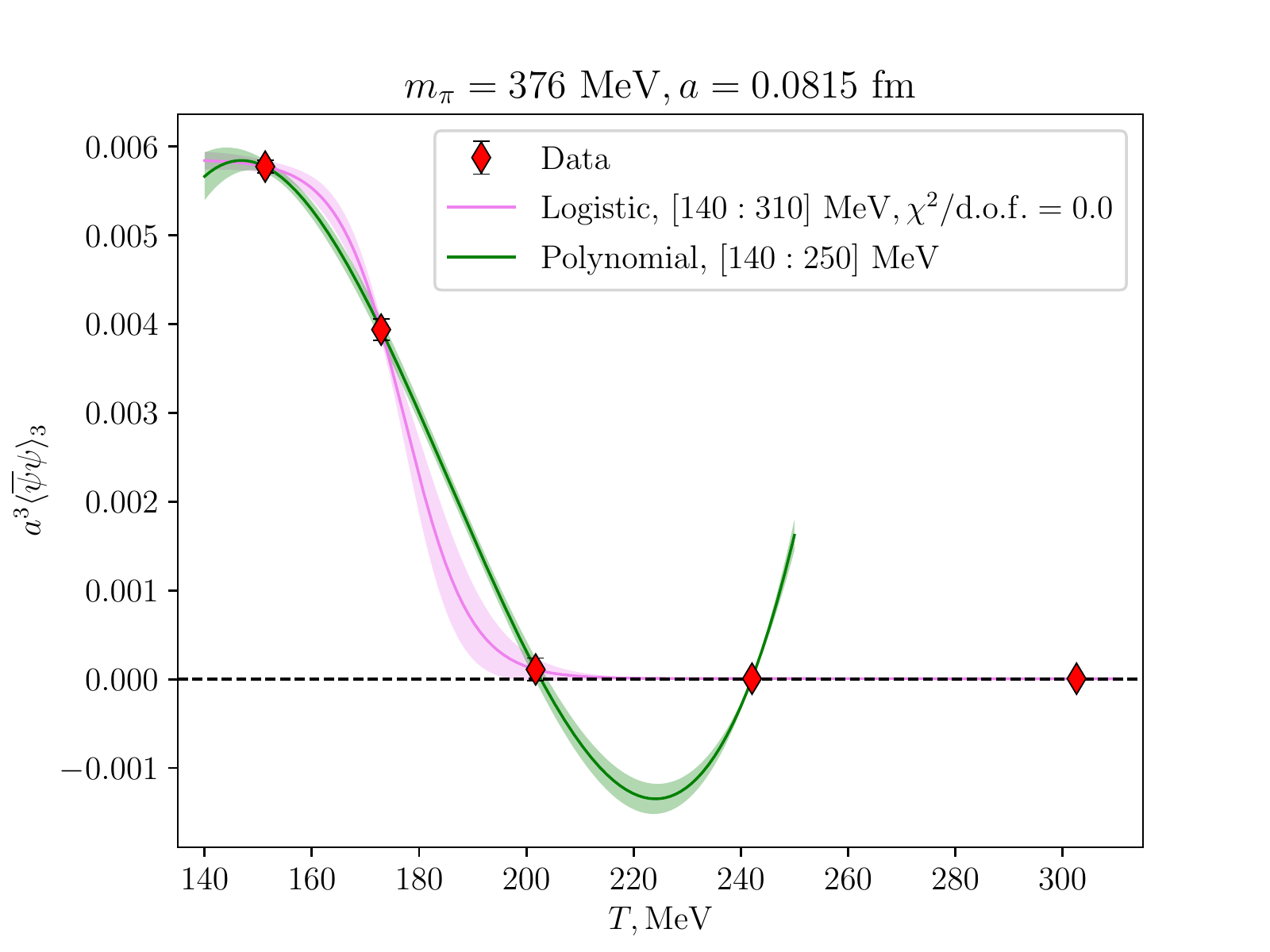}
    \caption{(Bare) $\langle \bar\psi\psi \rangle_3$ as a function of temperature $T$, for ensembles as indicated, and fits superimposed; see text for details.}
    \label{fig:subcondensate}
\end{figure*}

\begin{figure}[thb]
    \includegraphics[width=9cm]{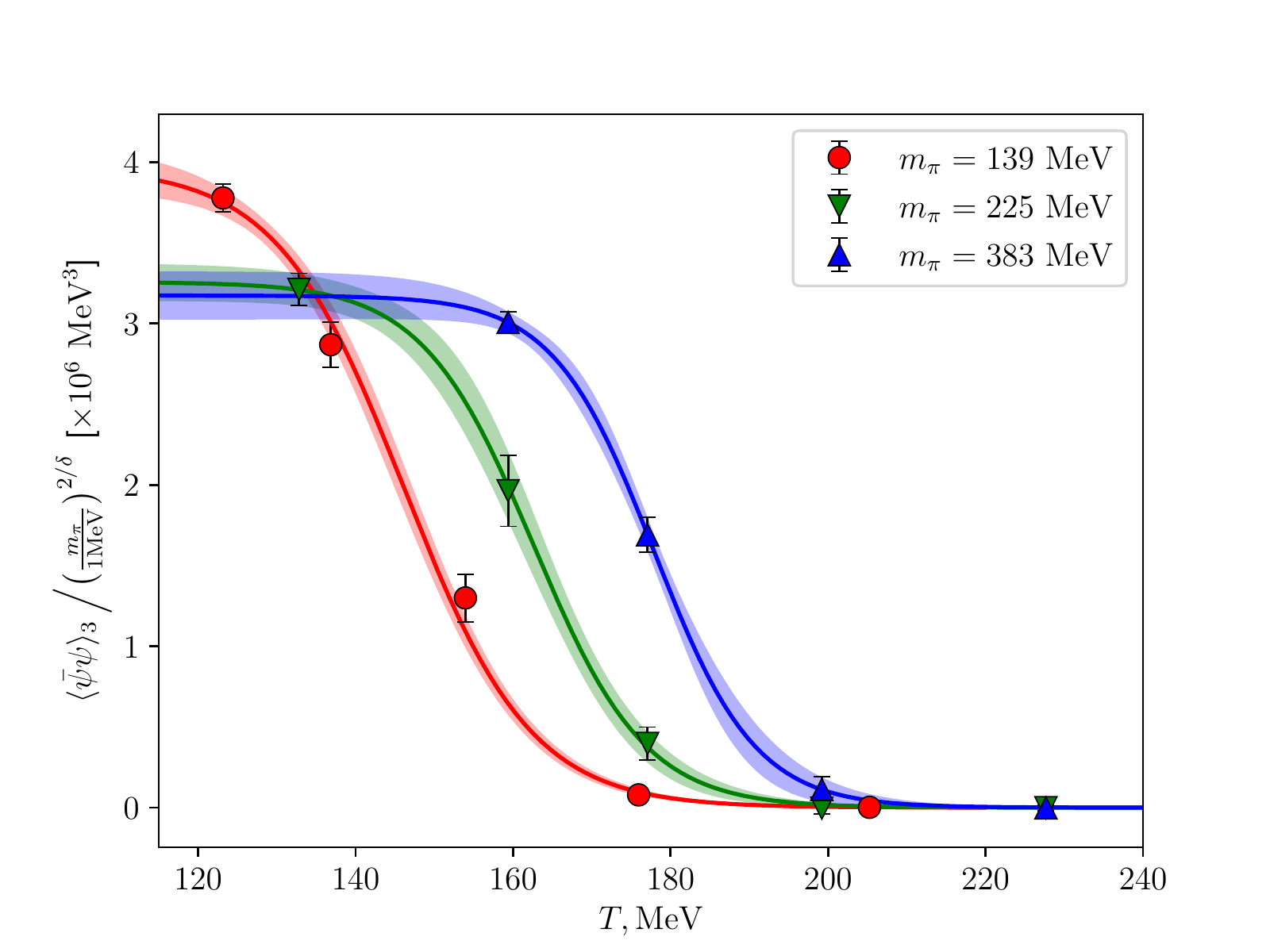}
    \caption{Search for the $O(4)$ scaling of $\langle\bar \psi \psi\rangle_3$: at the critical temperature,  $\frac{\langle\bar \psi \psi\rangle_3}{m_\pi^{2/\delta}}$ does not depend on the quark mass. The crossing point of the results for the two lightest masses picks a candidate critical temperature $T_0^{\delta} = 138(2)$ MeV.}
    \label{fig:crossinglogistic}
\end{figure}

In this work we present new results for the physical pion mass, together with an extended analysis of the results for higher pion masses \cite{Burger:2018fvb}. Simulations are performed in the fixed scale approach, where we keep the bare lattice parameters fixed and vary the temperature by varying the number of lattice points in the temporal direction $N_t$.
Gauge fields ensembles used in this paper are summarized in Table~\ref{table:sim}. New simulations
for the physical pion mass use Wilson-clover twisted mass fermions, with the parameters of the action corresponding to the zero temperature ensemble cB211.072.064 of ETMC \cite{Alexandrou:2018egz}. The number of configurations generated for these ensembles is given in Table~\ref{tab:stat}.
First results of our simulations with physical pion mass were presented in \cite{Kotov:2020hzm}.

The chiral condensate is an
(approximate) order parameter for the chiral symmetry in the light quark sector: 
\begin{equation}
    \langle \bar{\psi}\psi\rangle=\langle \bar{u}u\rangle+\langle \bar{d}d\rangle=\frac{T}{V}\frac{\partial Z}{\partial m_l}=\frac{1}{N_tN_s^3}\langle\tr M^{-1}\rangle.
    \label{eq:condensate}
\end{equation}
In numerical simulations the trace of $M^{-1}$ was measured using noisy stochastic estimator with 24 random volume sources. 
It is important to note that the bare chiral condensate (\ref{eq:condensate}) should be renormalized. However, in the fixed scale approach, applied in this paper, both additive and multiplicative renormalization factors are equal for all points. Thus, the renormalization procedure has no effect on the pseudo-critical temperatures extracted from the peak of $\chi_\Delta$, or, equivalently,  from the inflection point of the chiral condensate.

In Fig.~\ref{fig:condensate} we present the dependence of the chiral condensate~(\ref{eq:condensate}) on the temperature for the physical
pion mass. We extract the pseudo-critical temperature $T_{\Delta}$ from the inflection point of this dependence. For this purpose we fitted the chiral condensate with several functions, varying fitting interval:
\begin{itemize}
    \item Logistic: $A+B\tanh\frac{T-T_{\Delta}}{\delta T_{\Delta}}$,   \item $A+B\frac{T-T_{\Delta}}{\sqrt{\delta T^2+(T-T_{\Delta})^2}}$,
    \item Polynomial: $\Delta=a_{\Delta}+b_{\Delta}T+c_{\Delta}T^2+d_{\Delta}T^3$.
 \end{itemize}
We find that the quality of the fit worsens
when the upper limit approaches $300$ MeV. The final estimation is obtained by averaging over three functional forms and two fitting intervals [120:210] and [120:250] MeV (apart from the polynomial fit, which describes the data poorly at large interval [120:250] MeV). The difference between various functions/intervals was used to estimate the systematic uncertainty.
The final results for the pseudo-critical temperature $T_{\Delta}$ are quoted in Table~\ref{tab:crit_temp}. The results for  heavier pion masses are from our previous analysis \cite{Burger:2018fvb}. 

The chiral susceptibility, which is defined as the mass derivative of the chiral condensate $\chi_L=\frac{\partial}{\partial m_l}\langle \bar{\psi}\psi\rangle$, consists of connected $\chi_{\text{conn}}$ and disconnected  $\chi_{\text{disc}}$ contributions:
\begin{equation}
\begin{split}
    \chi_L=\frac{\partial}{\partial m_l}\langle \bar{\psi}\psi\rangle=\chi_{\text{disc}}+\chi_{\text{conn}},\\
    \chi_{\text{disc}}=\frac{T}{V}\left(\langle(\tr M^{-1})^2\rangle-\langle\tr M^{-1}\rangle^2\right),\\
    \chi_{\text{conn}}=-\frac{T}{V}\langle\tr M^{-2}\rangle.
\label{eq:sus}
\end{split}\end{equation}
Also, the  chiral susceptibility suffers from additive and multiplicative renormalizations, which, again, are not affecting the estimate of the pseudo-critical point.

In Fig.~\ref{fig:sus} we show the  chiral susceptibility 
for all ensembles. We note that this observable has a strong regular contribution, in addition to an additive renormalization. These features are qualitatively
clear in the plots: rather than the simple symmetric shape predicted by the EoS, the curves are skewed and have a long high temperature tail. For this reason, and given a small number of points in the transition region, we decided to use a model
independent estimate of the pseudo-critical temperature, based on cubic spline interpolation, instead of using an explicit functional form. To estimate the statistical uncertainty we added random Gaussian noise to each point, with the amplitude given by the statistical uncertainty of our data points. Typically we used $O(2000)$ splines for error estimate, and we find this a rather robust procedure. The dispersion of the different results can be appreciated from the plots, and we show the statistical errors in a few selected cases - the others behave similarly. 

Finally we compute  $\langle\bar \psi \psi\rangle_3$ as in  (\ref{eq:pbp3}).
As discussed above in Section 2,
$\langle\bar \psi \psi\rangle_3$
is free from linear additive renormalization as well as from linear correction to scaling.
It still needs a multiplicative renormalization, which is obviously the same $Z_p$ as for the chiral condensate,
Table~\ref{table:sim}. 

The temperature dependence of  $\langle\bar\psi\psi\rangle_3$ is
shown in Fig.~\ref{fig:subcondensate}.  We used several functional forms to fit the temperature dependence of this observable, which 
shows a fast fall-off with temperature. 
We then find that  the logistic curve is able to capture
the behaviour for a sizeable range, while polynomial fits are limited to a few points. Upon the fit we tried to keep the fit interval fixed in units $T/T_c$: for logistic fit the interval is roughly $[0.8, 1.5]$, for polynomial $[0.8, 1.25]$. We used both functional forms to estimate the pseudo-critical temperature and its systematic uncertainty. The final results for the pseudo-critical temperature extracted from the chiral susceptibility and for $\langle\bar\psi\psi\rangle_3$   for all ensembles are  presented in Table~\ref{tab:crit_temp}.
\section{The magnetic Equation of State and \texorpdfstring{$\langle\bar\psi\psi\rangle_3$}{pp3}} 
In Fig.~\ref{fig:crossinglogistic} we show the results for $\langle\bar\psi\psi\rangle_3$ as a function of temperature, 
with superimposed logistic fits. The results are converted to physical units using the lattice spacing and the multiplicative renormalization given in Table~\ref{table:sim} and divided by  $m_\pi^{2/\delta}$, with $\delta$ fixed at the $O(4)$ value. At the critical temperature $\langle \bar \psi \psi \rangle_3 \propto  m_\pi^{2/\delta}$, hence 
$\frac{\langle \bar \psi \psi \rangle_3}{  m_\pi^{2/\delta}}$ should not depend on the mass at the critical point. The logic is similar to the one adopted in
Refs.~\cite{Ding:2020xlj,Kocic:1992is} when searching for
the fixed point of $\chi_L/\chi_T$.
The crossing point of the curves for the lightest masses identifies 
a candidate critical temperature in the chiral limit, $T_0^{\delta} = 138(2)$ MeV, where the superscript $\delta$ refers to $T_0$ being estimated from the crossing point.
Possible scaling violations would result in a mass dependence of $T_0^{\delta}$. 

\begin{figure*}
     \includegraphics[width=9cm]{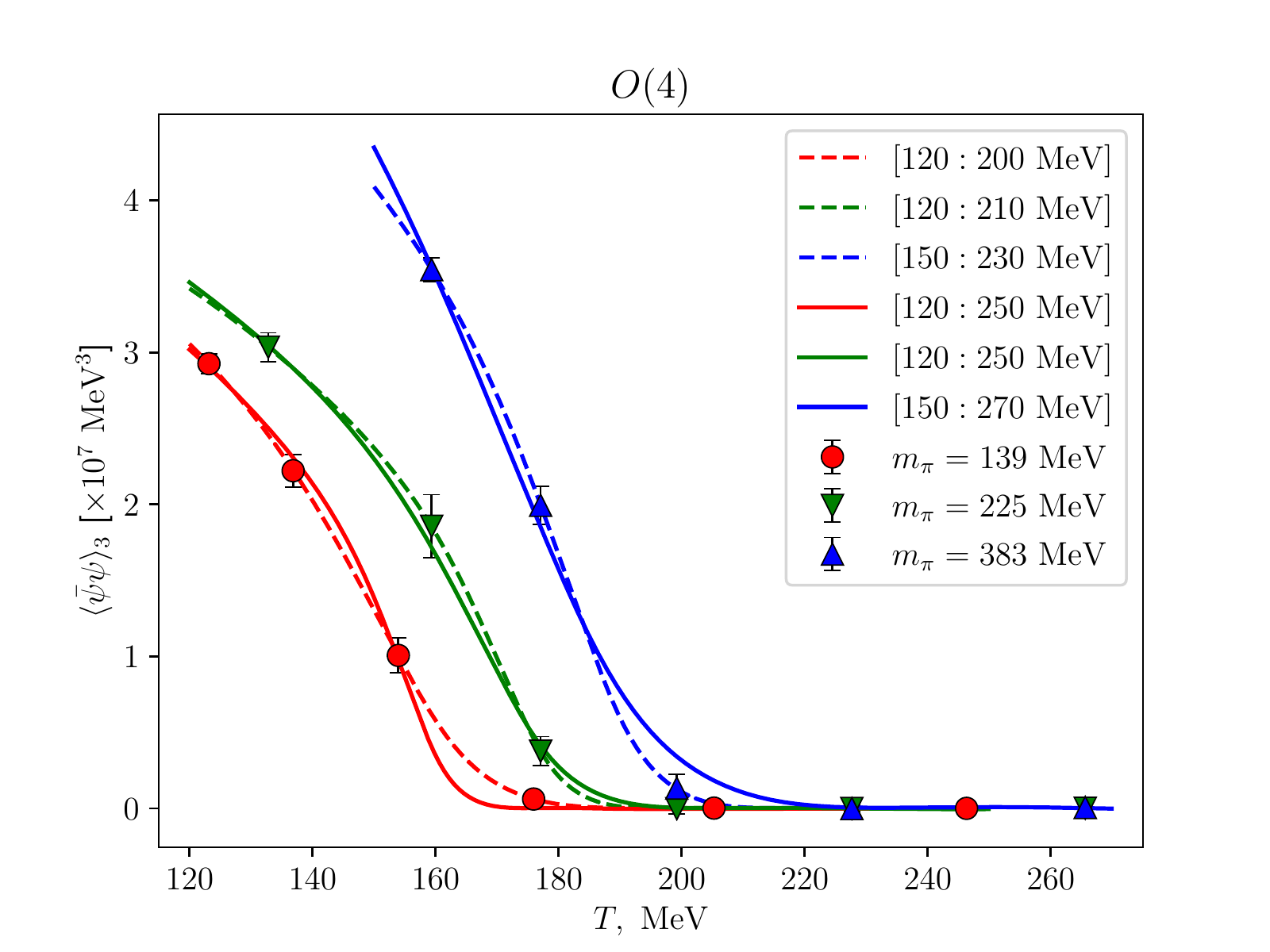}~~~~\includegraphics[width=9cm]{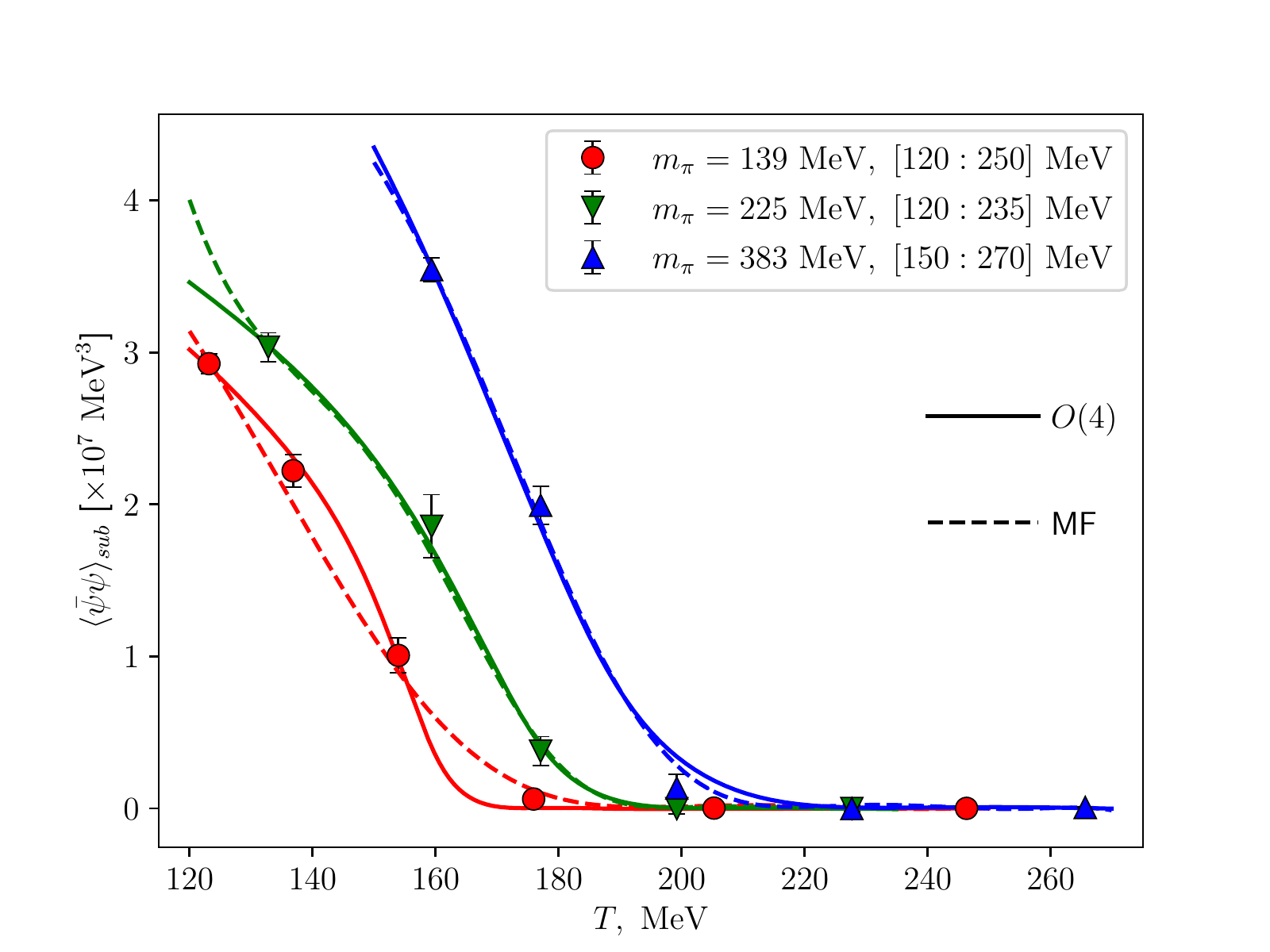}
\caption{Fits to the Equation of State: 3D $O(4)$ (left), 3D $O(4)$ and mean field (right). 
For $m_\pi = 139$ MeV the critical temperature $T_0^\text{EoS} = 142(2)$ MeV is consistent with that estimated from the chiral  extrapolation of the pseudo-critical temperatures (left);
the data are compatible with mean field, with a mild tension building up for the physical pion mass (right). }
\label{fig:physEoS}
\end{figure*}

\begin{figure}[thb]
    \includegraphics[width=9cm]{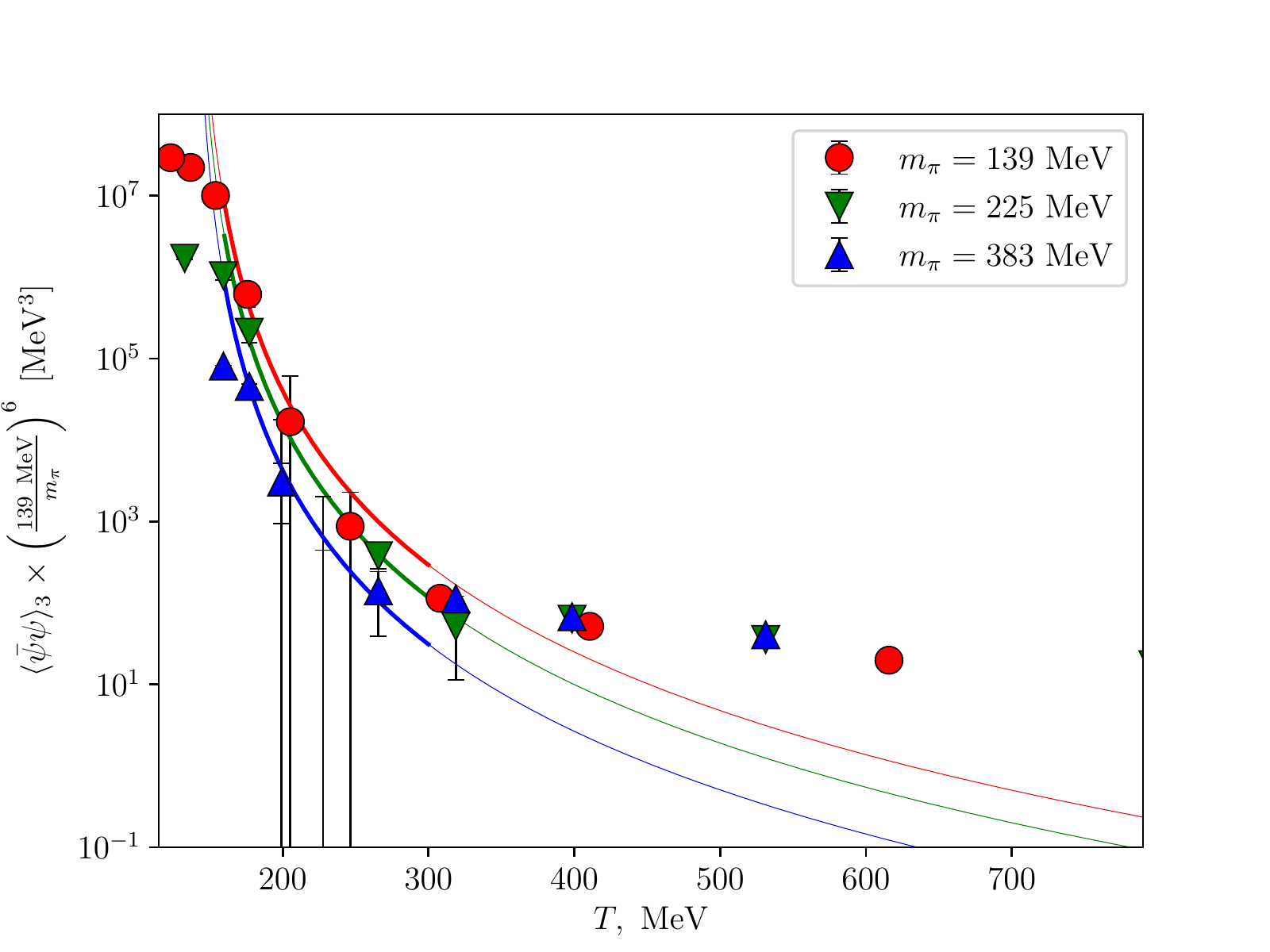}
    \caption{Fits to a constrained 3D $O(4)$ behaviour for the different pion masses:  the results
in the interval of temperatures [160:300]~MeV (marked bold) fare nicely through the data. For $T > 300$ MeV the behaviour is distinctly different, and  the data  follows  $m_l^3 \simeq m_\pi^6$, the anticipated high temperature leading behaviour. $\chi^2/\text{d.o.f.}$ of the fit is $0.5/0.7/0.7$ for pion masses $m_{\pi}=139$, $225$ and $383$ MeV correspondingly.}
    \label{fig:hight}
\end{figure}

\begin{figure}[tbh]
    \centering
    \includegraphics[width=9cm]{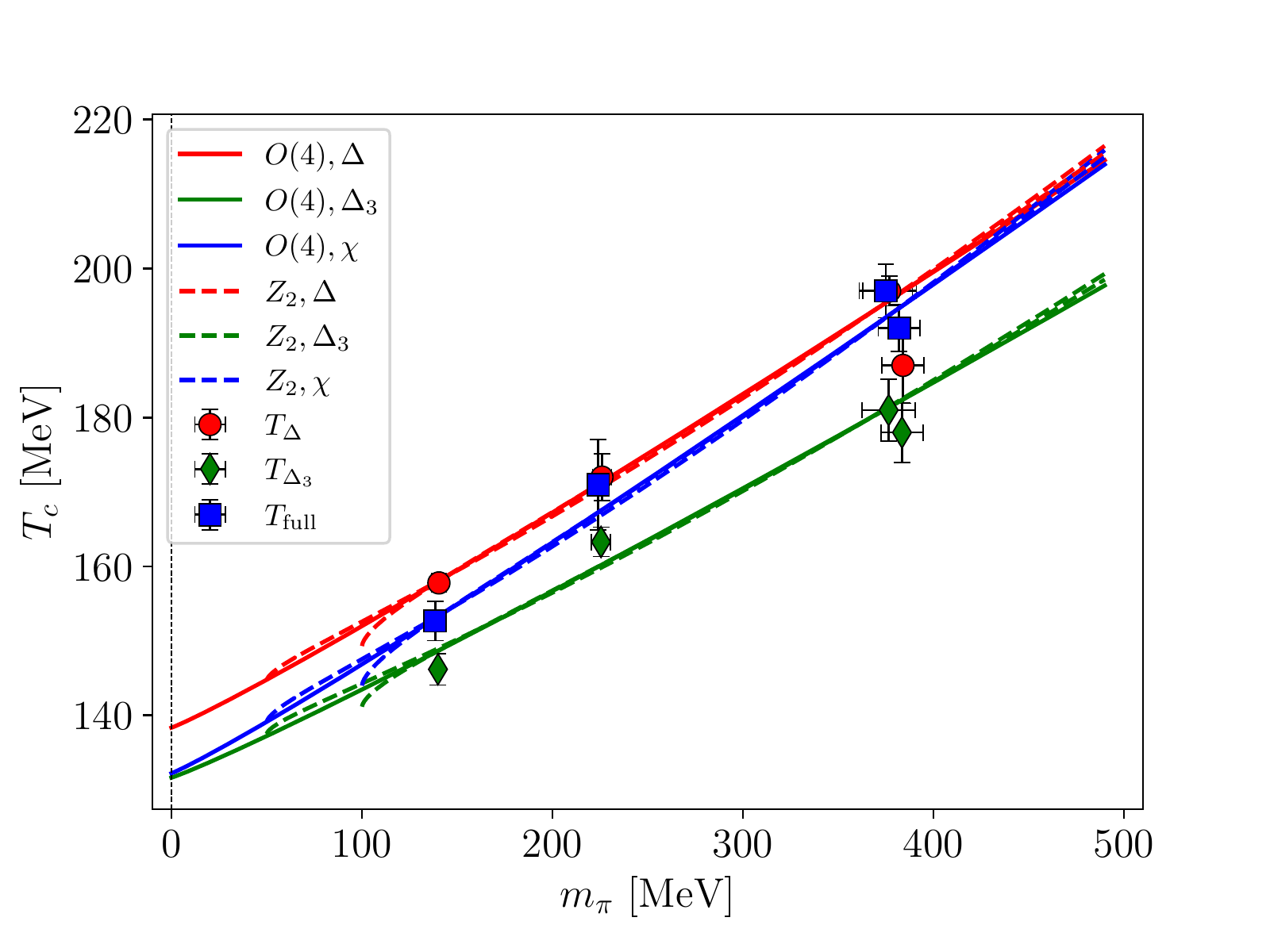}
    \caption{Pseudocritical temperatures from the chiral observables with superimposed the 3D $O(4)$ and $Z_2$ scaling fits described in the text. $\chi^2/\text{d.o.f.}$ for $O(4)$ fit is $\chi^2/\text{d.o.f.}=2.1/1.0/2.5$ for the chiral condensate $\Delta$, for the chiral susceptibility $\chi$ and the new observable $\Delta_3$. For $Z_2$ fits the $\chi^2/\text{d.o.f.}$ differs by maximum 20\% from $\chi^2/\text{d.o.f.}$ of $O(4)$ fit.}
    \label{fig:crittempvspion_z2_o4}
\end{figure}

Next, we fit to the 3D $O(4)$ EoS with open critical temperature. 
We rely on a smoothing spline to define the EoS and we introduce appropriate
(pion mass dependent) scaling parameters: $g_{\mathrm{fit}}(x) = a g\left(\frac{x - T_0^{\mathrm{EoS}}}{b}\right) $, where $g(x)$ is given by EoS for the subtracted condensate in Eq.~(\ref{eq:eosthree}).  In Fig.~\ref{fig:physEoS}, left, we show the results in physical units, with $g_{\mathrm{fit}}(x)$ superimposed as interpolating lines, which at a first sight look satisfactory.  However the would-be critical temperatures $T_0^{\mathrm{EoS}}$  are not constant with mass,  signaling residual scaling violations: we find  $T_0^{\mathrm{EoS}} = 142(2)$, 159(3), 174(2) MeV,
from light to heavy masses. Only $T_0^{\mathrm{EoS}}$ for the physical pion mass is compatible with the previous estimate $T_0^{\delta} = 138(2)$ MeV, but there are obvious violations for larger masses.  
In the right-hand side of Fig.~\ref{fig:physEoS}  we compare the mean field fits with the $O(4)$ EoS: as it was already clear from Fig.~\ref{fig:O4_pbp}, they are very close to each other. In the smaller interval they basically coincide,  only for the larger interval shown here,   and for the lightest mass,  $O(4)$ fits may be slightly favored. 

Finally,  in Fig.~\ref{fig:hight} we show high temperature fit of the condensate $\langle\bar{\psi}\psi\rangle_3$, constrained to the $O(4)$ behaviour: $\langle\bar{\psi}\psi\rangle_3\propto t^{-\gamma-2\beta\delta}$ for 
$T_0=138$ MeV\footnote{The sensitivity to $T_0$ is very moderate here. $T_0$ itself is rather poorly constrained by the fit. For example, if one keeps $T_0$ as a free parameter, then the fit for the physical pion mass $m_{\pi}=139$ MeV gives $T_0=132\pm 4$ MeV in the interval $[150:300]$~MeV and $T_0=145.7\pm0.6$ MeV in the interval $[160:340]$~MeV.}. The results are rescaled by $m_l^3 \simeq m_\pi^6$,  the anticipated high temperature leading behaviour. 
In the interval of temperatures $160$ MeV $ < T <  300$~MeV (marked bold) the $O(4)$ prediction fares nicely through the data. For $T \gtrsim 300$~MeV the behaviour is distinctly different, and the data collapse on a single curve, as expected of the leading mass term  according to Griffith analyticity, regardless the critical behaviour. This suggests that the temperature extent of the scaling window above $T_0$ extends up to about $300$~MeV. Interestingly, in a previous study \cite{Burger:2018fvb} we have found that this is also the threshold for a dilute instanton gas behaviour.

\section{The scaling of the pseudo-critical temperatures and the chiral limit}
In the summary Table~\ref{tab:crit_temp} we present all pseudocritical temperatures.
The results on the $B$ ensembles help monitoring the finite spacing effects, and we confirm that they are small also for the new
observables introduced here. A more detailed discussions  on the continuum limit  can be found in Ref. \cite{Burger:2018fvb}
as well as in ETMC papers, see e.g. Ref. \cite{Alexandrou:2021gqw}, and references therein. 

\begin{table}
\begin{center}
\begin{tabular}{|c|c|c|c|}
\hline
Ensemble & $T_\Delta$ & $T_{\Delta_3}$ & $T_{\chi}$ \\
\hline
M140& 157.8(7)(10)  & 146.2(21)(1) & 152.7(13)(23) \\
D210& 172(3)(1) & 163.3(18)(8) & 171(6)(1) \\
D370& 187(5)(1) & 178(4)(0) & 192(3)(1) \\
B370& 197(2)(0) & 181(1)(4) & 197(2)(3) \\
\hline
\end{tabular}
\caption{Pseudo-critical temperature extracted from
the chiral observables.}
\label{tab:crit_temp}
\end{center}
\end{table}

\begin{figure*}
    \centering
    \vskip -1cm
    \includegraphics[width=12cm]{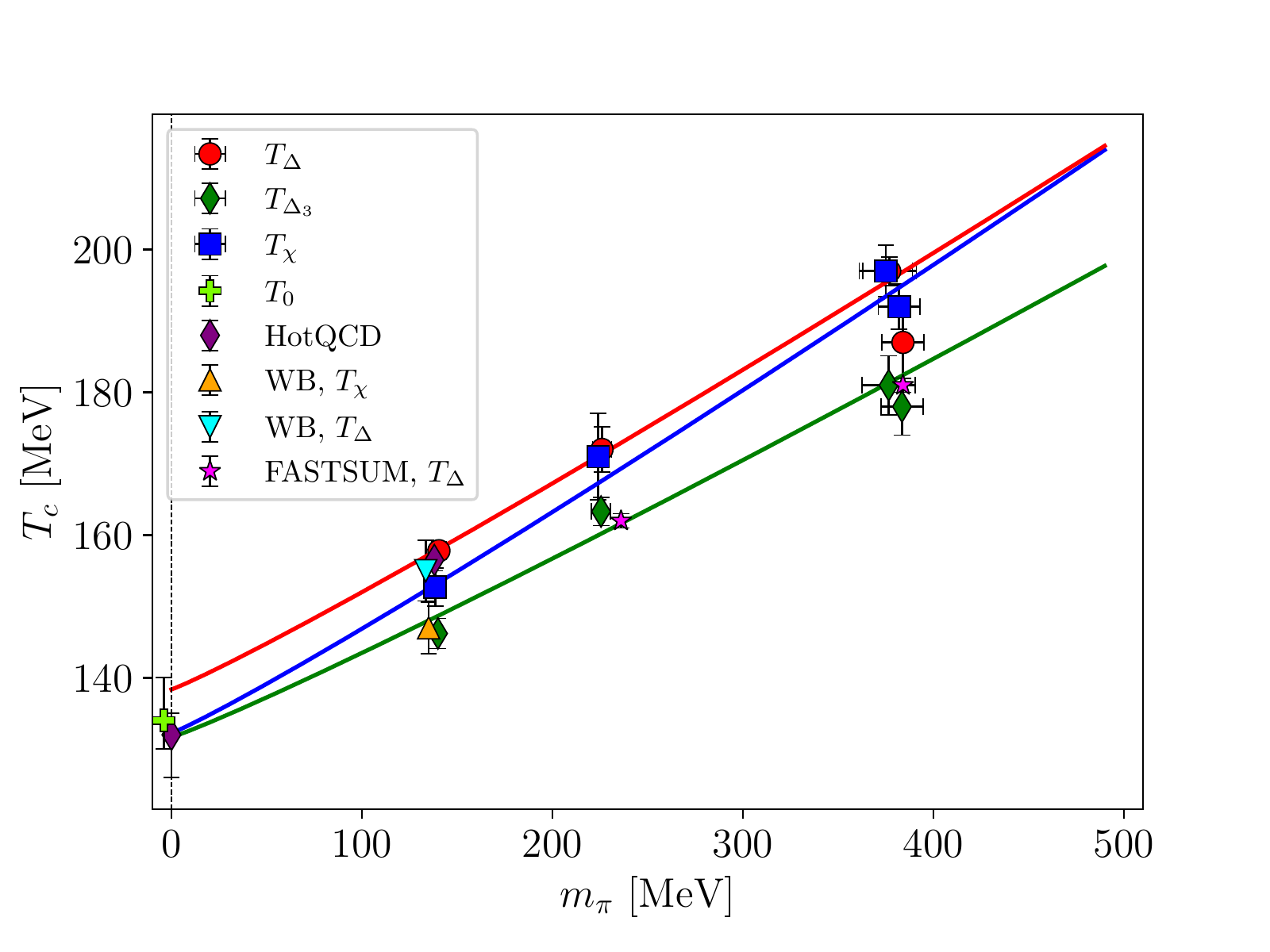}
    \caption{Pseudo-critical temperatures with their chiral extrapolations: comparison with the results from the HotQCD Collaboration~\cite{Bazavov:2018mes}, FASTSUM Collaboration~\cite{Aarts:2019hrg,Aarts:2020vyb}, Wuppertal-Budapest Collaboration~\cite{Borsanyi_2020}. The purple diamond at $m_{\pi}=0$ marks the
    critical temperature~\cite{Ding:2019prx} which compares well with our result $T_0 = 134^{+6}_{-4}$~MeV (light-green cross, slightly shifted for better readability).}
    \label{fig:crittempvspion2}
\end{figure*}

\begin{figure}[thb]
    \centering
    \includegraphics[width=9cm]{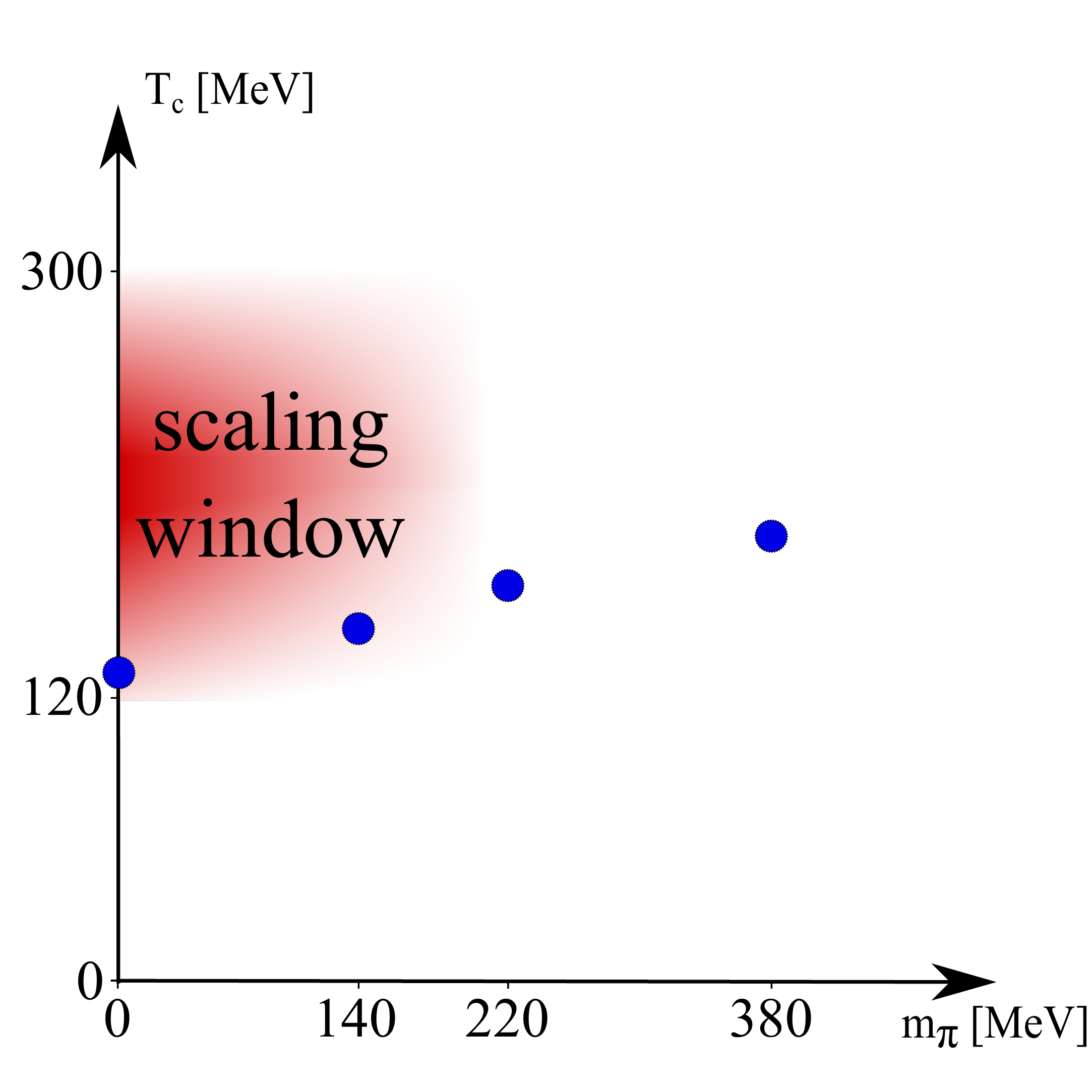}
    \caption{Sketch of the critical region: the critical temperature in the chiral limit $T_0$ is constrained from above from the pseudocritical temperature $T_{\Delta_3} =146(2)$ MeV. The averaged extrapolation  $T_0 = 134^{+6}_{-4}$~MeV satisfies this bound, and it is robust against different choices of critical scaling. It compares well with the estimate from anomalous scaling $T_0^\delta=138(2)$ MeV. The different red shades indicate the compatibility with $O(4)$ scaling. The labels indicate the pion masses that we have investigated, for temperatures ranging from 120 MeV till 500 MeV.}
    \label{fig:window}
\end{figure}

As discussed, we fit the pseudo-critical temperatures to a power law in the pion mass, see Eq.~\eqref{eq:tcmpi},
where $z_p$ is tabulated for the different observables 
in Table~\ref{tab:o4fit_temp}, $A$ is a universal constant
depending on the normalization scales, and $T_0 \equiv T_c(m_{\pi} \to 0)$ corresponds to the critical temperature in the chiral limit. We kept $2/(\beta\delta)=1.083$ fixed to the value given by $O(4)$ critical exponents. In Table \ref{tab:o4fit_temp} we present $T_0$ and ratio of $z_p$ extracted from the fit as well as the prediction of $O(4)$ scaling. 
There is a semi-quantitative agreement with $O(4)$ when considering the relative slopes of $\langle \bar \psi \psi \rangle$ and $\langle \bar \psi \psi \rangle_3$. When considering only the results for physical pion mass, with a critical temperature in the chiral limit within an acceptable range, $z_p/z_p(\bar{\psi}\psi_3)$ = 1.46, with a large error.

\begin{table}[h]
    \centering
    \begin{tabular}{|c|c|c|c|c|}
    \hline

    {\small Observable} & $T_0$ [MeV] & $z_p/z_{\bar{\psi}\psi_3}$ & $z_p/z_{\bar{\psi}\psi_3}$ {\small $O(4)$}  & $z_p$ {\small $O(4)$} \\

    \hline
    $\chi$ & 132(4) &  1.24(17) & 2.45(4) & 1.35(3)\\
    $\langle\bar{\psi}\psi\rangle$  & 138(2)     &  1.15(24) &  1.35(7) & 0.74(4) \\
    $\langle \bar{\psi}\psi\rangle_3$ & 132(3)  & 1 & 1 & 0.55(1)\\
    \hline
    \end{tabular}
    \caption{Critical temperature in the chiral limit $T_0$ and the ratio of $z_p$ extracted from the fit by Eq.~(\ref{eq:tcmpi}). Last two columns contain predictions for $z_p$ from 3D $O(4)$ scaling. Errors are only statistical.}
    \label{tab:o4fit_temp}
\end{table}

The average of the extrapolated values of  $T_c$ in the chiral limit
\begin{equation}
    T_0\equiv T_c(m_{\pi}\to 0) = 134^{+6}_{-4}~\text{MeV} 
    \label{eq:extrapolatedT}
\end{equation}
is compatible with the estimated from the crossing point $T_0^{\delta} = 138(2)$ MeV, 
and only slightly below the estimate from the EoS for physical pion mass.  This last point 
is consistent with the physical pion mass being close to the scaling window. Note that the error in Eq.~(\ref{eq:extrapolatedT}) contains also systematic uncertainty, coming from extrapolation of different observables. 

Similarly to Ref.~\cite{Burger:2011zc} we also consider  the breaking pattern corresponding to an effective restoration of the axial symmetry
$U(2)_L\!\times\!U(2)_R \to U(2)_V$, leading either to a second order transition with $\delta = 4.3(1), \beta = 0.40(4)$  or of 
first order transition, in which case the
pseudo-critical behaviour would be driven by an endpoint in the $Z_2$ universality class. In the first case the exponents are  close to the $O(4)$ ones and within our errors the results are indistinguishable.  For the first order scenario we rely on the experience gathered with three-flavor QCD, where it was found that the mixing associated with the critical endpoint
could well be low \cite{Karsch:2003va}. Hence, as in Ref.~\cite{Burger:2011zc}, we ignore the mixing and  fit to 
\begin{equation}
T_c(m_\pi) = T_0 + B (m_\pi^2 - m_c^2)^{1/{\beta \delta}}
\end{equation}
with $1/{\beta \delta} = 0.64$ for the $Z_2$ 3D universality class~\cite{Vicari:2008jw}. 
Also in this case, the critical exponents for $O(4)$ and $Z_2$ behaviour are close to each other,  and it is very difficult to distinguish between these two scenarios. One can clearly see from Fig.~\ref{fig:crittempvspion_z2_o4} that the data can be also described with $Z_2$ behaviour and critical mass up to $m_c\sim 100$ MeV. If, artificially, we constrained $m_c=0$, the lines would be 
indistinguishable from $O(4)$. All in all, we concur with other observations~\cite{Ding:2020xlj,Ding:2020rtq} that extrapolation alone does not suffice to discriminate among different scenarios. A possible way to distinguish between different universal behaviours would include simulations with lower pion masses~\cite{Ding:2019prx}.
On the positive side, the extrapolated values of the pseudo-critical temperatures $T$ are robust (up to the critical pion mass $m^\pi_\text{cr}$) against the choice of critical exponents.

In Fig.~\ref{fig:crittempvspion2} we compare our estimations of the pseudo-critical temperature with the results obtained by staggered~\cite{Bazavov:2018mes,Borsanyi_2020} and Wilson~\cite{Aarts:2019hrg,Aarts:2020vyb} fermions. The nice agreement of our results with others can be appreciated. We also superimpose $O(4)$ fits to our pseudo-critical temperatures and present our estimation of $T_0$ in the chiral limit which compares well with the estimations of HotQCD Collaboration~\cite{Ding:2019prx}.

\section{Summary}
The chiral behaviour of the QCD transition is a difficult, much studied
problem. Direct tests of universality are hampered by scaling violations,
and by the similarity among different scenarios. Here we have tackled
these issues by introducing an order parameter, whose behaviour is closer
to the critical one, hence with a reduced contribution from scaling violating terms; and by exploring a larger parameter region, with the goal
to follow the evolution from a scaling behaviour till a regular one,
unrelated with criticality. 

Particularly interesting is the behaviour with physical quark masses. We have measured the pseudo-critical temperature for three different observables, finding critical temperatures in agreement with staggered
estimates for the chiral condensate and the chiral susceptibility. The 
new order parameter $\langle \bar \psi \psi \rangle_3$
gives a lower pseudo-critical temperature $T_{\Delta_3} = 146.2(21)(1)$~MeV, in accordance with its being closer
to the true critical behaviour. 
We have used three different strategies for the computation of the critical temperature in the chiral limit: the first one is based on the scaling of  $\langle \bar \psi \psi \rangle_3$ at the critical temperature $T_0(m_{\pi}=0)$, the second one is based on the fit of $\langle \bar \psi \psi \rangle_3$ to the magnetic Equation of State, which should give the value of $T_0$ for each single pion mass, barring scaling violations, the last one is based on extrapolation of the pseudo-critical temperatures extracted from different observables to the chiral limit. A search for the anomalous scaling of $\langle \bar \psi \psi \rangle_3$ at the chiral transition picks the critical temperature
of $T_0 = 138(2)$ MeV. For all the masses the results  are apparently well described by the universal EoS, however only for the physical pion mass the estimated critical temperature $T_0 = 142(2)$ MeV approaches the results obtained by extrapolation,  
$T_0 = 134^{+6}_{-4}$ MeV. We note that the pseudo-critical temperature $T_{\Delta_3} = 146.2(21)(1)$ MeV for a physical pion mass serves as an upper bound for the critical temperature. As a final result for the critical
temperature we quote $T_0 = 134^{+6}_{-4}$~MeV. In Fig.~\ref{fig:crittempvspion2} we present our results for the pseudo-critical temperatures, their chiral extrapolation and comparison to the results of other groups, with which we find a nice agreement.

For larger masses the 3D $O(4)$ EoS fits are
indistinguishable from mean field, while a mild tension appears for the physical mass, with $O(4)$ slightly favoured. The high temperature behaviour shows a clear threshold at temperatures about $300$ MeV, above which a trend
consistent with $O(4)$ scaling gives way to a simple leading 
order analytic scaling dictated by Griffith analyticity.
As a side comment,  in a previous study \cite{Burger:2018fvb} we have found that this is also the threshold for a dilute instanton gas behaviour. A threshold around the same temperature was also observed in other
lattice studies\cite{Rohrhofer:2019qwq,Alexandru:2019gdm,Cardinali:2021mfh}. Our observations  are compatible with an onset of scaling 
around the physical value of the pion mass, and temperatures below $300$ MeV. In Fig.~\ref{fig:window} we sketch the critical region, highlighting the compatibility with $O(4)$ scaling. Our results seem to be in contradiction with the results of~\cite{Braun:2010vd,Braun:2020ada}, where
scaling is observed only for very tiny pion masses. Possible sources of this discrepancy and connection to the presented results are the subject of future work and will be discussed elsewhere.

We also studied the possibility of first order phase transition in the chiral limit, in which case the observables are expected to scale according to $Z_2$ universality class with some critical pion mass $m_c$. Our data can be also described with $Z_2$ behaviour and critical mass up to $m_c\sim 100$~MeV. Thus we conclude  that extrapolation alone does not suffice to discriminate among different scenarios. Similar conclusions were reached in the  two flavor model~\cite{Burger:2011zc}. A possible way to distinguish between different universal behaviour would include simulations with lower pion masses~\cite{Ding:2019prx,Ding:2020xlj}, hoping to observe directly a first order transition.  While a direct observation of such a first order transition would settle the issue, a lack of observation makes it  difficult to exclude that such a transition appears at very low masses. We hope that by exploiting more the universal properties of the Equation of State, possibly using some of the strategies outlined here, and  in comparison with analytic approaches, may provide a more solid answer.

It would also be interesting to repeat a similar analysis using staggered results for pion masses smaller than physical \cite{Ding:2020xlj} as well as with FRG results \cite{Braun:2020ada}. Also, Fig.~\ref{fig:O4_pbp} suggests that results below the critical temperature in the chiral limit may be useful to further discriminate true critical behaviour from mean field.

 This work should be ameliorated and extended along several lines:  we have not performed  a continuum limit extrapolation. The indirect checks performed by the ETMC
collaboration as well as by ourselves \cite{Burger:2018fvb}, and  the good consistency of the results at the larger mass, see Table~\ref{tab:crit_temp},
give some confidence that residual spacing corrections should not exceed a few percent. The fixed scale approach has several advantages, but
we have to rely heavily on interpolations. Results on finer lattices should also help in this respect, by producing results at intermediate temperatures.  Since the ETM Collaboration has recently released a new
set of zero temperature tuned parameters~\cite{Alexandrou:2021gqw}, we hope to return to this point in the future. 
Finally, we have completed an analysis of correlators and screening masses on the configurations for physical pion mass, 
to complement the discussion on chiral and axial symmetry, and the results will appear soon~\cite{progress}. 

 \section*{Acknowledgements}
We thank Frithjof Karsch, Anirban Lahiri, Jan Pawlowski and Wolfgang Unger for valuable discussions and correspondence. This work uses the ETMC public code, and we are grateful to the ETM Collaboration, in particular to Roberto Frezzotti, Karl Jansen and Carsten Urbach, for helpful 
discussions on several aspects of the twisted mass formulation. This work is partially supported by  STRONG-2020, a 
European Union’s Horizon 2020 research and innovation programme under grant agreement No. 824093.
The work of A.Yu.K. and A.T. was supported by RFBR grant 18-02-40126. A.T. acknowledges support from the "BASIS" foundation.
Numerical simulations have been carried out using computing resources of CINECA (based on the agreement between INFN and CINECA, on the ISCRA project IsB20), the supercomputer of Joint Institute for Nuclear Research ``Govorun'', and the computing resources of the federal collective usage center Complex for Simulation and Data Processing for Mega-science Facilities at NRC ``Kurchatov Institute'',~\url{http://ckp.nrcki.ru/}.


\begin{thebibliography}{10}

\bibitem{Ding:2020rtq}
Heng-Tong Ding.
\newblock {New developments in lattice QCD on equilibrium physics and phase
  diagram}.
\newblock In {\em {28th International Conference on Ultrarelativistic
  Nucleus-Nucleus Collisions}}, 2 2020, 2002.11957.

\bibitem{Guenther:2021lrv}
Jana~N. Guenther.
\newblock {Overview of the QCD phase diagram: Recent progress from the
  lattice}.
\newblock {\em Eur. Phys. J. A}, 57(4):136, 2021.

\bibitem{Kotov:2020hzm}
Andrey~{\relax Yu}. Kotov, Maria~Paola Lombardo, and Anton~M. Trunin.
\newblock {Finite temperature QCD with $N_f=2+1+1$ Wilson twisted mass fermions
  at physical pion, strange and charm masses}.
\newblock {\em Eur. Phys. J.}, A56(8):203, 2020, 2004.07122.

\bibitem{Kotov:2019dby}
Andrey~Yu. Kotov, Maria~Paola Lombardo, and Anton~M. Trunin.
\newblock {Fate of the $\eta^{'}$ in the quark gluon plasma}.
\newblock {\em Phys. Lett. B}, 794:83--88, 2019, 1903.05633.

\bibitem{Burger:2018fvb}
Florian Burger, Ernst-Michael Ilgenfritz, Maria~Paola Lombardo, and Anton
  Trunin.
\newblock {Chiral observables and topology in hot QCD with two families of
  quarks}.
\newblock {\em Phys. Rev. D}, 98(9):094501, 2018, 1805.06001.

\bibitem{Burger:2017xkz}
Florian Burger, Ernst-Michael Ilgenfritz, Maria~Paola Lombardo, Michael
  Müller-Preussker, and Anton Trunin.
\newblock {Topology (and axion's properties) from lattice QCD with a dynamical
  charm}.
\newblock {\em Nucl. Phys. A}, 967:880--883, 2017, 1705.01847.

\bibitem{Burger:2011zc}
Florian Burger, Ernst-Michael Ilgenfritz, Malik Kirchner, Maria~Paola Lombardo,
  Michael Müller-Preussker, Owe Philipsen, Carsten Urbach, and Lars
  Zeidlewicz.
\newblock {Thermal QCD transition with two flavors of twisted mass fermions}.
\newblock {\em Phys. Rev. D}, 87(7):074508, 2013, 1102.4530.

\bibitem{Pisarski:1983ms}
Robert~D. Pisarski and Frank Wilczek.
\newblock {Remarks on the Chiral Phase Transition in Chromodynamics}.
\newblock {\em Phys. Rev. D}, 29:338--341, 1984.

\bibitem{Rajagopal:1992qz}
Krishna Rajagopal and Frank Wilczek.
\newblock {Static and dynamic critical phenomena at a second order QCD phase
  transition}.
\newblock {\em Nucl. Phys. B}, 399:395--425, 1993, hep-ph/9210253.

\bibitem{Pelissetto:2013hqa}
Andrea Pelissetto and Ettore Vicari.
\newblock {Relevance of the axial anomaly at the finite-temperature chiral
  transition in QCD}.
\newblock {\em Phys. Rev. D}, 88(10):105018, 2013, 1309.5446.

\bibitem{Kawarabayashi:1980dp}
Ken Kawarabayashi and Nobuyoshi Ohta.
\newblock {The Problem of $\eta$ in the Large $N$ Limit: Effective Lagrangian
  Approach}.
\newblock {\em Nucl. Phys.}, B175:477--492, 1980.

\bibitem{Kawarabayashi:1980uh}
Ken Kawarabayashi and Nobuyoshi Ohta.
\newblock {On the Partial Conservation of the U(1) Current}.
\newblock {\em Prog. Theor. Phys.}, 66:1789, 1981.

\bibitem{Nicola:2019ohb}
Angel Gómez~Nicola, Jacobo Ruiz De~Elvira, and Andrea Vioque-Rodríguez.
\newblock {The QCD topological charge and its thermal dependence: the role of
  the $\eta'$}.
\newblock {\em JHEP}, 11:086, 2019, 1907.11734.

\bibitem{Horvatic:2018ztu}
Davor Horvati\'c, Dalibor Kekez, and Dubravko Klabu\v~car.
\newblock {$\eta'$ and $\eta$ mesons at high T when the $U_A$(1) and chiral
  symmetry breaking are tied}.
\newblock {\em Phys. Rev. D}, 99(1):014007, 2019, 1809.00379.

\bibitem{Kapusta:2019ktm}
Joseph~I. Kapusta, Ermal Rrapaj, and Serge Rudaz.
\newblock {Is Hyperon Polarization in Relativistic Heavy Ion Collisions
  Connected to Axial U(1) Symmetry Breaking at High Temperature?}
\newblock {\em Phys. Rev. C}, 101(3):031901, 2020, 1910.12759.

\bibitem{Kapusta:1995ww}
Joseph~I. Kapusta, D.~Kharzeev, and Larry~D. McLerran.
\newblock {The Return of the prodigal Goldstone boson}.
\newblock {\em Phys. Rev. D}, 53:5028--5033, 1996, hep-ph/9507343.

\bibitem{Shuryak:1993ee}
Edward~V. Shuryak.
\newblock {Which chiral symmetry is restored in hot QCD?}
\newblock {\em Comments Nucl. Part. Phys.}, 21(4):235--248, 1994,
  hep-ph/9310253.

\bibitem{Resch:2017vjs}
Simon Resch, Fabian Rennecke, and Bernd-Jochen Schaefer.
\newblock {Mass sensitivity of the three-flavor chiral phase transition}.
\newblock {\em Phys. Rev. D}, 99(7):076005, 2019, 1712.07961.

\bibitem{Schaefer:2013isa}
Bernd-Jochen Schaefer and Mario Mitter.
\newblock {Three-flavor chiral phase transition and axial symmetry breaking
  with the functional renormalization group}.
\newblock {\em Acta Phys. Polon. Supp.}, 7(1):81--90, 2014, 1312.3850.

\bibitem{Gao:2020qsj}
Fei Gao and Jan~M. Pawlowski.
\newblock {QCD phase structure from functional methods}.
\newblock {\em Phys. Rev. D}, 102(3):034027, 2020, 2002.07500.

\bibitem{Braun:2020ada}
Jens Braun, Wei-jie Fu, Jan~M. Pawlowski, Fabian Rennecke, Daniel Rosenbl\"uh,
  and Shi Yin.
\newblock {Chiral susceptibility in ( 2+1 )-flavor QCD}.
\newblock {\em Phys. Rev. D}, 102(5):056010, 2020, 2003.13112.

\bibitem{Mazur:2018pjw}
Lukas Mazur, Olaf Kaczmarek, Edwin Laermann, and Sayantan Sharma.
\newblock {The fate of axial U(1) in 2+1 flavor QCD towards the chiral limit}.
\newblock {\em PoS}, LATTICE2018:153, 2019, 1811.08222.

\bibitem{Sharma:2018syt}
Sayantan Sharma.
\newblock {The fate of $U_A(1)$ and topological features of QCD at finite
  temperature}.
\newblock In {\em {11th International Workshop on Critical Point and Onset of
  Deconfinement}}, 1 2018, 1801.08500.

\bibitem{Fukaya:2017wfq}
Hidenori Fukaya.
\newblock {Can axial U(1) anomaly disappear at high temperature?}
\newblock {\em EPJ Web Conf.}, 175:01012, 2018, 1712.05536.

\bibitem{Sharma:2017yjc}
Sayantan Sharma.
\newblock {The fate of U$_A$(1) and topological features of QCD at finite
  temperature}.
\newblock {\em PoS}, CPOD2017:086, 2018.

\bibitem{Schmidt:2017bjt}
Christian Schmidt and Sayantan Sharma.
\newblock {The phase structure of QCD}.
\newblock {\em J. Phys. G}, 44(10):104002, 2017, 1701.04707.

\bibitem{Tomiya:2016jwr}
A.~Tomiya, G.~Cossu, S.~Aoki, H.~Fukaya, S.~Hashimoto, T.~Kaneko, and J.~Noaki.
\newblock {Evidence of effective axial U(1) symmetry restoration at high
  temperature QCD}.
\newblock {\em Phys. Rev. D}, 96(3):034509, 2017, 1612.01908.
\newblock [Addendum: Phys.Rev.D 96, 079902 (2017)].

\bibitem{Aarts:2019hrg}
Gert Aarts et~al.
\newblock {Spectral quantities in thermal QCD: a progress report from the
  FASTSUM collaboration}.
\newblock In {\em {37th International Symposium on Lattice Field Theory}},
  2019, 1912.09827.

\bibitem{Aarts:2020vyb}
G.~Aarts et~al.
\newblock {Properties of the QCD thermal transition with $N_f=2+1$ flavours of
  Wilson quark}.
\newblock {\em arXiv:2007.04188}, 2020.

\bibitem{Ding:2019prx}
H.T. Ding et~al.
\newblock {Chiral Phase Transition Temperature in ( 2+1 )-Flavor QCD}.
\newblock {\em Phys. Rev. Lett.}, 123(6):062002, 2019, 1903.04801.

\bibitem{Umeda:2016qdo}
T.~Umeda, S.~Ejiri, R.~Iwami, K.~Kanaya, H.~Ohno, A.~Uji, N.~Wakabayashi, and
  S.~Yoshida.
\newblock {O(4) scaling analysis in two-flavor QCD at finite temperature and
  density with improved Wilson quarks}.
\newblock {\em PoS}, LATTICE2016:376, 2017, 1612.09449.

\bibitem{Ejiri:2009ac}
S.~Ejiri, F.~Karsch, E.~Laermann, C.~Miao, S.~Mukherjee, P.~Petreczky,
  C.~Schmidt, W.~Soeldner, and W.~Unger.
\newblock {On the magnetic equation of state in (2+1)-flavor QCD}.
\newblock {\em Phys. Rev. D}, 80:094505, 2009, 0909.5122.

\bibitem{Ding:2020xlj}
H.~T. Ding, S.~T. Li, Swagato Mukherjee, A.~Tomiya, X.~D. Wang, and Y.~Zhang.
\newblock {Correlated Dirac Eigenvalues and Axial Anomaly in Chiral Symmetric
  QCD}.
\newblock {\em Phys. Rev. Lett.}, 126(8):082001, 2021, 2010.14836.

\bibitem{Kaczmarek:2021ser}
Olaf Kaczmarek, Lukas Mazur, and Sayantan Sharma.
\newblock {Eigenvalue spectra of QCD and the fate of $U_A(1)$ breaking towards
  the chiral limit}.
\newblock 2 2021, 2102.06136.

\bibitem{Kaczmarek:2020sif}
Olaf Kaczmarek, Frithjof Karsch, Anirban Lahiri, Lukas Mazur, and Christian
  Schmidt.
\newblock {QCD phase transition in the chiral limit}.
\newblock 3 2020, 2003.07920.

\bibitem{Aoki:2021qws}
S.~Aoki, Y.~Aoki, H.~Fukaya, S.~Hashimoto, C.~Rohrhofer, and K.~Suzuki.
\newblock {Role of axial U(1) anomaly in chiral susceptibility of QCD at high
  temperature}.
\newblock 3 2021, 2103.05954.

\bibitem{Aoki:2020noz}
S.~Aoki, Y.~Aoki, G.~Cossu, H.~Fukaya, S.~Hashimoto, T.~Kaneko, C.~Rohrhofer,
  and K.~Suzuki.
\newblock {Study of the axial $U(1)$ anomaly at high temperature with lattice
  chiral fermions}.
\newblock {\em Phys. Rev. D}, 103(7):074506, 2021, 2011.01499.

\bibitem{Buchoff:2013nra}
Michael~I. Buchoff, Michael Cheng, Norman~H. Christ, H.-T. Ding, Chulwoo Jung,
  F.~Karsch, Zhongjie Lin, R.~D. Mawhinney, Swagato Mukherjee, P.~Petreczky,
  Dwight Renfrew, Chris Schroeder, P.~M. Vranas, and Hantao Yin.
\newblock Qcd chiral transition, $u(1{)}_{A}$ symmetry and the dirac spectrum
  using domain wall fermions.
\newblock {\em Phys. Rev. D}, 89:054514, Mar 2014, 1309.4149.

\bibitem{Suzuki:2020rla}
Kei Suzuki, Sinya Aoki, Yasumichi Aoki, Guido Cossu, Hidenori Fukaya, Shoji
  Hashimoto, and Christian Rohrhofer.
\newblock {Axial U(1) symmetry and mesonic correlators at high temperature in
  $N_f=2$ lattice QCD}.
\newblock In {\em {37th International Symposium on Lattice Field Theory}}, 1
  2020, 2001.07962.

\bibitem{Kanazawa:2015xna}
Takuya Kanazawa and Naoki Yamamoto.
\newblock {U (1) axial symmetry and Dirac spectra in QCD at high temperature}.
\newblock {\em JHEP}, 01:141, 2016, 1508.02416.

\bibitem{Aoki:2012yj}
Sinya Aoki, Hidenori Fukaya, and Yusuke Taniguchi.
\newblock {Chiral symmetry restoration, eigenvalue density of Dirac operator
  and axial U(1) anomaly at finite temperature}.
\newblock {\em Phys. Rev. D}, 86:114512, 2012, 1209.2061.

\bibitem{Brandt:2019ksy}
Bastian~B. Brandt, Marco Cè, Anthony Francis, Tim Harris, Harvey~B. Meyer,
  Hartmut Wittig, and Owe Philipsen.
\newblock {Testing the strength of the $\text{U}_A(1)$ anomaly at the chiral
  phase transition in two-flavour QCD}.
\newblock {\em PoS}, CD2018:055, 2019, 1904.02384.

\bibitem{Brandt:2016daq}
Bastian~B. Brandt, Anthony Francis, Harvey~B. Meyer, Owe Philipsen, Daniel
  Robaina, and Hartmut Wittig.
\newblock {On the strength of the $U_A(1)$ anomaly at the chiral phase
  transition in $N_f=2$ QCD}.
\newblock {\em JHEP}, 12:158, 2016, 1608.06882.

\bibitem{Cossu:2013uua}
Guido Cossu, Sinya Aoki, Hidenori Fukaya, Shoji Hashimoto, Takashi Kaneko,
  Hideo Matsufuru, and Jun-Ichi Noaki.
\newblock {Finite temperature study of the axial U(1) symmetry on the lattice
  with overlap fermion formulation}.
\newblock {\em Phys. Rev. D}, 87(11):114514, 2013, 1304.6145.
\newblock [Erratum: Phys.Rev.D 88, 019901 (2013)].

\bibitem{Chiu:2013wwa}
Ting-Wai Chiu, Wen-Ping Chen, Yu-Chih Chen, Han-Yi Chou, and Tung-Han Hsieh.
\newblock {Chiral symmetry and axial U(1) symmetry in finite temperature QCD
  with domain-wall fermion}.
\newblock {\em PoS}, LATTICE2013:165, 2014, 1311.6220.

\bibitem{Braun:2010vd}
Jens Braun, Bertram Klein, and Piotr Piasecki.
\newblock {On the scaling behavior of the chiral phase transition in QCD in
  finite and infinite volume}.
\newblock {\em Eur. Phys. J. C}, 71:1576, 2011, 1008.2155.

\bibitem{Kotov:2021hri}
Andrey~Yu Kotov, Maria~Paola Lombardo, and Anton Trunin.
\newblock {Gliding Down the QCD Transition Line, from $N_f$ = 2 till the Onset
  of Conformality}.
\newblock {\em Symmetry}, 13(10):1833, 2021.

\bibitem{Unger:2010wcq}
Wolfgang Unger.
\newblock {\em {The chiral phase transition of QCD with 2+1 flavors : a lattice
  study on Goldstone modes and universal scaling}}.
\newblock PhD thesis, U. Bielefeld (main), 2010.

\bibitem{Engels:1999wf}
Jurgen Engels and Tereza Mendes.
\newblock {Goldstone mode effects and scaling function for the
  three-dimensional O(4) model}.
\newblock {\em Nucl. Phys.}, B572:289--304, 2000, hep-lat/9911028.

\bibitem{Engels:2011km}
J.~Engels and F.~Karsch.
\newblock {The scaling functions of the free energy density and its derivatives
  for the 3d O(4) model}.
\newblock {\em Phys. Rev.}, D85:094506, 2012, 1105.0584.

\bibitem{Engels:2003nq}
J.~Engels, L.~Fromme, and M.~Seniuch.
\newblock {Correlation lengths and scaling functions in the three-dimensional
  O(4) model}.
\newblock {\em Nucl. Phys.}, B675:533--554, 2003, hep-lat/0307032.

\bibitem{Kocic:1992is}
Aleksandar Kocic, John~B. Kogut, and Maria-Paola Lombardo.
\newblock {Universal properties of chiral symmetry breaking}.
\newblock {\em Nucl. Phys.}, B398:376--404, 1993, hep-lat/9209010.

\bibitem{Werner:2019hxc}
Markus Werner et~al.
\newblock {Hadron-Hadron Interactions from $N_f=2+1+1$ Lattice QCD: The
  $\rho$-resonance}.
\newblock {\em Eur. Phys. J.}, A56(2):61, 2020, 1907.01237.

\bibitem{Carrasco:2014cwa}
N.~Carrasco et~al.
\newblock {Up, down, strange and charm quark masses with N$_f$ = 2+1+1 twisted
  mass lattice QCD}.
\newblock {\em Nucl. Phys.}, B887:19--68, 2014, 1403.4504.

\bibitem{Alexandrou:2020okk}
C.~Alexandrou et~al.
\newblock {Nucleon axial and pseudoscalar form factors from lattice QCD at the
  physical point}.
\newblock {\em Phys. Rev. D}, 103(3):034509, 2021, 2011.13342.

\bibitem{Alexandrou:2018egz}
Constantia Alexandrou et~al.
\newblock {Simulating twisted mass fermions at physical light, strange and
  charm quark masses}.
\newblock {\em Phys. Rev.}, D98(5):054518, 2018, 1807.00495.

\bibitem{Alexandrou:2021gqw}
C.~Alexandrou et~al.
\newblock {Quark masses using twisted mass fermion gauge ensembles}.
\newblock 4 2021, 2104.13408.

\bibitem{Bazavov:2018mes}
A.~Bazavov et~al.
\newblock {Chiral crossover in QCD at zero and non-zero chemical potentials}.
\newblock {\em Phys. Lett.}, B795:15--21, 2019, 1812.08235.

\bibitem{Borsanyi_2020}
Szabolcs Borsanyi, Zoltan Fodor, Jana~N. Guenther, Ruben Kara, Sandor~D. Katz,
  Paolo Parotto, Attila Pasztor, Claudia Ratti, and Kálman~K. Szabó.
\newblock Qcd crossover at finite chemical potential from lattice simulations.
\newblock {\em Physical Review Letters}, 125(5), Jul 2020.

\bibitem{Karsch:2003va}
F.~Karsch, C.~R. Allton, S.~Ejiri, S.~J. Hands, O.~Kaczmarek, E.~Laermann, and
  C.~Schmidt.
\newblock {Where is the chiral critical point in three flavor QCD?}
\newblock {\em Nucl. Phys. B Proc. Suppl.}, 129:614--616, 2004,
  hep-lat/0309116.

\bibitem{Vicari:2008jw}
Ettore Vicari and Haralambos Panagopoulos.
\newblock {Theta dependence of SU(N) gauge theories in the presence of a
  topological term}.
\newblock {\em Phys. Rept.}, 470:93--150, 2009, 0803.1593.

\bibitem{Rohrhofer:2019qwq}
C.~Rohrhofer, Y.~Aoki, G.~Cossu, H.~Fukaya, C.~Gattringer, L.~Ya. Glozman,
  S.~Hashimoto, C.~B. Lang, and S.~Prelovsek.
\newblock {Symmetries of spatial meson correlators in high temperature QCD}.
\newblock {\em Phys. Rev. D}, 100(1):014502, 2019, 1902.03191.

\bibitem{Alexandru:2019gdm}
Andrei Alexandru and Ivan Horv\'ath.
\newblock {Possible New Phase of Thermal QCD}.
\newblock {\em Phys. Rev. D}, 100(9):094507, 2019, 1906.08047.

\bibitem{Cardinali:2021mfh}
Marco Cardinali, Massimo D'Elia, and Andrea Pasqui.
\newblock {Thermal monopole condensation in QCD with physical quark masses}.
\newblock 7 2021, 2107.02745.

\bibitem{progress}
Andrey~Yu. Kotov, Maria~Paola Lombardo, and Anton Trunin.
\newblock {Work in progress}.

\end{thebibliography}
\end{document}